\documentclass[a4paper]{article}
\pdfoutput=1
\usepackage[hyphens]{url}
\usepackage[top=2cm,bottom=3cm,left=2.4cm,right=2.4cm]{geometry}
\usepackage{mathtools,amssymb}
\usepackage[usenames,dvipsnames,table]{xcolor}
\usepackage[titletoc]{appendix}
\usepackage{mathtools,amssymb}
\usepackage[makeroom]{cancel}
\usepackage{graphicx}
\usepackage{epstopdf}
\usepackage{mathrsfs}
\usepackage{enumitem}
\usepackage{stackrel}
\usepackage{float}
\usepackage{booktabs}
\usepackage[normalem]{ulem}  
\usepackage{cancel} 
\usepackage{hyperref}
\usepackage[draft,color,final]{showkeys} 
\usepackage{longtable}
\usepackage{physics}
\usepackage{empheq}

\usepackage{cite} 
\usepackage{comment}
\usepackage{tikz}

\DeclareMathAlphabet{\mathpzc}{OT1}{pzc}{m}{it}

\allowdisplaybreaks
\allowbreak

\definecolor{refkey}{gray}{0.75}
\definecolor{labelkey}{RGB}{155,48,48}
\renewcommand*\showkeyslabelformat[1]{%
  \fbox{\parbox[t]{0.8\marginparwidth}{\raggedright\normalfont\scriptsize\url{#1}}}}

\usepackage{etoolbox}
\makeatletter
\patchcmd{\hyper@makecurrent}{%
    \ifx\Hy@param\Hy@chapterstring
    \let\Hy@param\Hy@chapapp
    \fi
}{%
    \iftoggle{inappendix}{
	\@checkappendixparam{chapter}%
	\@checkappendixparam{section}%
	\@checkappendixparam{subsection}%
	\@checkappendixparam{subsubsection}%
	\@checkappendixparam{paragraph}%
	\@checkappendixparam{subparagraph}%
    }{}%
}{}{ \errmessage{failed to patch}}

\newcommand*{\@checkappendixparam}[1]{%
	\def\@checkappendixparamtmp{#1}%
	\ifx\Hy@param\@checkappendixparamtmp
	\let\Hy@param\Hy@appendixstring
	\fi
}
\makeatletter

\newtoggle{inappendix}
\togglefalse{inappendix}

\apptocmd{\appendix}{\toggletrue{inappendix}}{}{\errmessage{failed to patch}}
\apptocmd{\subappendices}{\toggletrue{inappendix}}{}{\errmessage{failed to patch}}

\newcommand{\lsim}{\mathrel{\hbox{\rlap{\lower .55ex
\hbox{$\sim$}} \kern-.3em \raise.4ex \hbox{$<$}}}}
\newcommand{\gsim}{\mathrel{\hbox{\rlap{\lower.55ex
\hbox{$\sim$}} \kern-.3em \raise.4ex \hbox{$<$}}}}

\begin{document}


\newcommand{\partiald}[2]{\dfrac{\partial #1}{\partial #2}}
\newcommand{\be}{\begin{equation}}
\newcommand{\ee}{\end{equation}}
\newcommand{\f}{\frac}
\newcommand{\s}{\sqrt}
\newcommand{\lm}{\mathcal{L}}
\newcommand{\wm}{\mathcal{W}}
\newcommand{\om}{\mathcal{O}_{n}}
\newcommand{\bea}{\begin{eqnarray}}
\newcommand{\eea}{\end{eqnarray}}
\newcommand{\ba}{\begin{align}}
\newcommand{\ea}{\end{align}}
\newcommand{\ep}{\epsilon}

\def\gap#1{\vspace{#1 ex}}
\def\be{\begin{equation}}
\def\ee{\end{equation}}
\def\bal{\begin{array}{l}}
\def\ba#1{\begin{array}{#1}}  
\def\ea{\end{array}}
\def\bea{\begin{eqnarray}}
\def\eea{\end{eqnarray}}
\def\beas{\begin{eqnarray*}}
\def\eeas{\end{eqnarray*}}
\def\del{\partial}
\def\eq#1{(\ref{#1})}
\def\fig#1{Fig \ref{#1}} 
\def\re#1{{\bf #1}}
\def\bull{$\bullet$}
\def\nn{\nonumber}
\def\ub{\underbar}
\def\nl{\hfill\break}
\def\ni{\noindent}
\def\bibi{\bibitem}
\def\vev#1{\langle #1 \rangle} 
\def\mattwo#1#2#3#4{\left(\begin{array}{cc}#1&#2\\#3&#4\end{array}\right)} 
\def\tgen#1{T^{#1}}
\def\half{\frac12}
\def\floor#1{{\lfloor #1 \rfloor}}
\def\ceil#1{{\lceil #1 \rceil}}

\def\Tr{{\rm Tr}}

\def\mysec#1{\gap1\ni{\bf #1}\gap1}
\def\mycap#1{\begin{quote}{\footnotesize #1}\end{quote}}

\def\Red#1{{\color{red}#1}}
\def\blue#1{{\color{blue}#1}}
\def\Om{\Omega}
\def\a{\alpha}
\def\b{\beta}
\def\l{\lambda}
\def\g{\gamma}
\def\e{\epsilon}
\def\Si{\Sigma}
\def\p{\phi}
\def\z{\zeta}

\def\lan{\langle}
\def\ran{\rangle}

\def\bit{\begin{item}}
\def\eit{\end{item}}
\def\benu{\begin{enumerate}}
\def\eenu{\end{enumerate}}

\def\tr{{\rm tr}}
\def\intk#1{{\int\kern-#1pt}}


\parindent=0pt
\parskip = 10pt

\def\al{\alpha}
\def\ga{\gamma}
\def\Ga{\Gamma}
\def\G{\Gamma}
\def\be{\beta}
\def\de{\delta}
\def\De{\Delta}
\def\ep{\epsilon}
\def\ro{\rho}
\def\la{\lambda}
\def\La{\Lambda}
\def\ka{\kappa}
\def\om{\omega}
\def\si{\sigma}
\def\th{\theta}
\def\ze{\zeta}
\def\ne{\eta}
\def\del{\partial}
\def\cdev{\nabla}

\def\gh{\hat{g}}
\def\Rh{\hat{R}}
\def\Boxh{\hat{\Box}}
\def\Kb{\mathcal{K}}
\def\phit{\tilde{\phi}}
\def\gt{\tilde{g}}
\newcommand{\h}{\hat}
\newcommand{\ti}{\tilde}
\newcommand{\sD}{{\mathcal{D}}}
\newcommand{\colored}[1]{ {\color{turquoise} #1 } }
\newcommand{\propBbd}{\mathcal{G}}
\newcommand{\propBB}{\mathbb{G}}
\newcommand{\christof}[3]{ {\Ga^{#1}}_{#2 #3}}
\def\ads{AdS$_{\text{2}}$~}
\def\GN{G$_{\text{N}}$~}
\def\zb{{\bar{z}}}
\def\fb{\bar{f}}
\def\delb{\bar{\del}}
\def\wb{\bar{w}}
\def\gb{\bar{g}}
\def\gp{g_+}
\def\gm{g_-}
\def\phit{\tilde{\phi}}
\def\xb{\bar{x}}
\def\yb{\bar{y}}
\def\xp{x_+}
\def\xm{x_-}
\def\finv{\mathfrak{f}_i}
\def\fbinv{\bar{\mathfrak{f}}_i}
\def\gc{\mathfrak{g}}
\def\gcb{\bar{ \mathfrak{g}}}
\def\disc{\mathcal{D}}
\def\rhp{\mathbb{H}}
\def\picklemma{Schwarz-Pick lemma}
\def\mobius{M\"{o}bius~}
\def\ft{\tilde{f}}
\def\zet{\tilde{\ze}}
\def\taut{\tilde{\tau}}
\def\thet{\tilde{\theta}}
\def\slr{\ensuremath{\mathbb{SL}(2,\mathbb{R})}}
\def\slc{\ensuremath{\mathbb{SL}(2,\mathbb{C})}}
\def\nh{\hat{n}}
\def\cD{\mathcal{D}}
\def\Lfg{\mathfrak{f}}
\def\mO{\mathcal{O}}

\newcommand{\com}{\textcolor{blue}}
\newcommand{\new}[1]{{\color[rgb]{1.0,0.,0}#1}}
\newcommand{\old}[1]{{\color[rgb]{0.7,0,0.7}\sout{#1}}}

\renewcommand{\real}{\ensuremath{\mathbb{R}}}

\newcommand*{\Cdot}[1][1.25]{%
  \mathpalette{\CdotAux{#1}}\cdot%
}
\newdimen\CdotAxis
\newcommand*{\CdotAux}[3]{%
    {%
	\settoheight\CdotAxis{$#2\vcenter{}$}%
	\sbox0{%
	    \raisebox\CdotAxis{%
		\scalebox{#1}{%
		    \raisebox{-\CdotAxis}{%
			$\mathsurround=0pt #2#3$%
		    }%
		}%
	    }%
	}%
	\dp0=0pt %
	\sbox2{$#2\bullet$}%
	\ifdim\ht2<\ht0 %
	\ht0=\ht2 %
	\fi
	\sbox2{$\mathsurround=0pt #2#3$}%
	\hbox to \wd2{\hss\usebox{0}\hss}%
    }%
}

\newcommand\hcancel[2][black]{\setbox0=\hbox{$#2$}%
\rlap{\raisebox{.45\ht0}{\textcolor{#1}{\rule{\wd0}{1pt}}}}#2} 

\renewcommand{\arraystretch}{2.5}%
\renewcommand{\floatpagefraction}{.8}%

\def\newthing{\marginpar{{\color{red}****}}}
\reversemarginpar


\def\tu{\tau}
\def\ze{z}
\def\d{\partial}
\def\L{\varphi}  

\DeclareRobustCommand{\rchi}{{\mathpalette\irchi\relax}}
\newcommand{\irchi}[2]{\raisebox{\depth}{$#1\chi$}}


\hypersetup{pageanchor=false}
\begin{titlepage}
    \begin{flushright}	ICTS/2018/01, TIFR/TH/18-03
    \end{flushright}

    \vspace{.4cm}
    \begin{center}
	\noindent{\Large \bf{Holographic dual to charged SYK from 3D
            Gravity and Chern-Simons}}\\
	\vspace{1cm} Adwait
        Gaikwad$^a$\footnote{adwait@theory.tifr.res.in}, Lata
        Kh Joshi$^a$\footnote{latakj@theory.tifr.res.in}, Gautam
        Mandal$^a$\footnote{mandal@theory.tifr.res.in}, and Spenta
        R. Wadia$^b$\footnote{spenta.wadia@icts.res.in}

	\vspace{.5cm}
	\begin{center}
	    {\it a. Department of Theoretical Physics}\\
	    {\it Tata Institute of Fundamental Research, Mumbai 400005, 
	    India.}\\
	    \vspace{.5cm}
	    {\it b. International Centre for Theoretical Sciences}\\
	    {\it Tata Institute of Fundamental Research, Shivakote,
	    Bengaluru 560089, India.}
	\end{center}

	\gap2


    \end{center}

     \begin{abstract}
       In this paper, we obtain a bulk dual to the low energy sector of the SYK model, including SYK model with $U(1)$ charge, by Kaluza-Klein (KK) reduction from three dimensions.  We show that KK reduction of the 3D Einstein action plus its boundary term gives the Jackiw-Teitelboim (JT) model in 2D with the appropriate 1D boundary term. The size of the KK radius gets identified with the value of the dilaton in the resulting near-AdS$_2$ geometry. In presence of U(1) charge, the 3D model additionally includes a $U(1)$ Chern-Simons (CS) action. In order to describe a boundary theory with non-zero chemical potential, we also introduce a coupling between CS gauge field and bulk gravity.  The 3D CS action plus the new coupling term with appropriate boundary terms reduce in two dimensions to a BF-type action plus a source term and boundary terms. The KK reduced 2D theory represents the soft sector of the charged SYK model. The pseudo-Nambu-Goldstone modes of combined {\it Diff}/$\mathbb{SL}(2,\mathbb{R})$ and $U(1)_{\rm local}/U(1)$ transformations are represented by combined large diffeomorphisms and large gauge transformations. The effective action of the former is reproduced by the action cost of the latter in the bulk dual, after appropriate identification of parameters. We compute chaotic correlators from the bulk and reproduce the result that the contribution from the ``boundary photons'' corresponds to zero Liapunov exponent.

\end{abstract}

\gap5

\centerline{\it Dedicated to the memory of Joe Polchinski.}

\end{titlepage}

\pagenumbering{roman}
\tableofcontents
\pagenumbering{arabic}
\setcounter{page}{1}

\section{Introduction and Summary}\label{sec:intro}
Finding simple field theories dual to gravitational systems has been a long standing goal. In the 1990's matrix models provided an early impetus to such a search, where a main ingredient of the duality was symmetries. For example, the $c=1$ matrix model was found to be governed by $W_\infty$ symmetries; consequently the dynamics could be abstracted in terms of quantizing coadjoint  orbits of $W_\infty$. The latter had a natural realization in terms of two dimensional string theory, thus establishing a connection between $c=1$ matrix model and two dimensional string theory \cite{Dhar:1992hr}. A similar approach was taken recently in \cite{Mandal:2017thl}, based on coadjoint orbits of the Virasoro symmetry group of the SYK model \cite{Sachdev:1992fk, Sachdev:2010um, Kitaev-talks:2015, PhysRevX.5.041025}, to construct a bulk dual to its `soft sector'. In this paper, we explore the possibility of obtaining such bulk duals by Kaluza-Klein reduction from 3D theories\footnote{We thank E. Witten and D. Stanford for initial suggestions regarding   a possible connection of the viewpoint of \cite{Mandal:2017thl} to   three dimensions.}.

The SYK model, briefly, is a model of interacting fermions at a single point. Its relevance to the physics of black holes stems from the fact that the model saturates the quantum chaos bound. The main feature of the theory responsible for this is the appearance of time
reparametrization symmetry (henceforth called {\it
  Diff}) \footnote{{\it Diff} represents ${\it Diff}\ \mathbb{S}^1$ or ${\it Diff}\ \mathbb{R}^1$,
  depending on whether the system is at a finite temperature or zero
  temperature.} at the strong coupling (IR fixed point) in the large $N$ limit  (see the original
papers mentioned above as well as \cite{Polchinski:2016xgd,
  Maldacena:2016hyu, Gurau:2016lzk, Witten:2016iux, Klebanov:2016xxf}
for further developments). The symmetry is spontaneously broken to
\slr, leading to Nambu Goldstone (NG) modes parameterized by the
coset. The IR theory is singular since the NG modes have precisely
zero action (there is no analogue of the pion kinetic energy term). To
regulate the theory, one has to introduce a small breaking of the
reparameterization symmetry by being slightly away from the strong
coupling fixed point; the dynamics of the pseudo NG modes is governed
by an action described by the Schwarzian of the reparametrization
function (see \eq{schwarzian-intro} below).

The original SYK model involved Majorana fermions which did not carry
any charge.  In \cite{Davison:2016ngz}, a generalized SYK model with
Dirac fermions with a global $U(1)$ was introduced. At the strong
coupling fixed point, this $U(1)$ symmetry is enhanced to local $U(1)$
transformations\footnote{Spontaneously broken by the vacuum to
  $U(1)$.}. As in the uncharged case, the theory is singular at the
strong coupling limit. The dynamics of the combined pseudo NG modes,
denoted by $\varphi(\tau) \in ${\it Diff}/\slr, and $\exp[i \phi(\tau)] \in
U(1)_{\rm local}/U(1)$ is given by the following action \cite{Davison:2016ngz}:
\begin{align}
S[\L,\phi] =&\ S_1 + S_2~,
\label{total-s-eff-intro}\\
S_1 =&\ - \f{\g}{4\pi^2}\int_0^\b d\tau\{\tan(\pi \varphi(\tau)/\b), \tau\} 
\label{schwarzian-intro}~;
\\ S_2 =&\ \f{K}{2}\int_0^\b d\tau \left[\partial_\tau\phi -i \mu
  (\partial_\tau \ep(\tau)) \right]^2~, \;\; \ep(\tau) \equiv \varphi(\tau)-
\tau.
\label{sigma-model-intro}
\end{align}
In the above, $K$ and $\g$ are constants depending on the coupling $J$
and chemical potential $\mu$, and are $\sim 1/J$. We have used the
following notation for the Schwarzian of a function $f(x)$
\[
\{ f, x\} \equiv \f{f'''}{f'}- \f32 \left(\f{f''}{f'}\right)^{\!2}~,
\]
where $\prime$ denotes derivative with respect to $x$. The first term in \eq{total-s-eff-intro} is the familiar Schwarzian from the uncharged case, written at a finite temperature ~$1/\beta$. The second term describes the action for the new set of pseudo Nambu Goldstone modes corresponding to $U(1)$ gauge transformations and their coupling to the {\it Diff} modes.  The pseudo NG bosons represent the `soft' sector of the charged SYK model, which satisfies the following condition
\begin{align}
\om \ll  J , \;
\mu \ll J~,
\label{soft}
\end{align}
where $\om$ denotes the frequency corresponding to the relative time
$\tau_1-\tau_2$ of a bilocal variable $G(\tau_1, \tau_2)$. The configurations of the charged SYK model are schematically represented by the left panel of Figure \ref{fig-diff-gauge}.

The pseudo Nambu Goldstone modes of reparameterization invariance of the SYK model are reminiscent of large diffeomorphisms of an asymptotically \ads geometry. This idea has been implemented in the bulk dual, in the Jackiw-Teitelboim (JT) models \cite{Almheiri:2014cka, Maldacena:2016upp} as well as in the Polyakov gravity model \cite{Mandal:2017thl}, based on the coadjoint orbit point of view mentioned above. A precise form of asymptotic \ads geometries, which implements the large diffeomorphisms, appears in \cite{Mandal:2017thl} and is given by
\begin{align}
ds^2_{{\rm AAdS}_2} = \frac{l^2}{z^2}\left(dz^2+d\tau^2\left(1-
\f{z^2}{2}\bigg \{\tan\left(\f{\pi \L(\tau)}{\beta} \right),\tau\bigg\}\right)^{\!\! 2}\right)~,
\label{aads2intro}
\end{align}
where $\{\tau, z\}$ are the time and radial coordinates in the gravity
and $l$ denotes AdS length scale. As above, $\L(\tau)$ \footnote{At finite inverse temperature $\beta$ the Schwarzian mode is parameterized as $f(\tau) = \beta/\pi \tan(\pi\L(\tau)/\beta)$ where $f(\tau)$ is {\it Diff} $\mathbb{R}^1$ and $\L(\tau)$ is {\it Diff} $\mathbb{S}^1$ function.} denotes elements
of the {\it Diff} group. Note that from the bulk viewpoint, the
unbroken subgroup \slr \ is evident from \eq{aads2intro}, since the
Schwarzian is invariant under the \slr \ transformation
\begin{equation}
 \tan\left({\pi \L(\tau)}/{\beta} \right) \to \f{a \tan\left({\pi \L(\tau)}/{\beta} \right) + b }{c \tan\left({\pi \L(\tau)}/{\beta} \right) + d}, \quad \text{with }ad-bc=1. \nn
\end{equation}
As has been shown in \cite{Almheiri:2014cka, Maldacena:2016upp,
  Mandal:2017thl}, in the presence of appropriate symmetry-breaking
terms, the large diffeomorphisms of the near-\ads geometry well
capture the low energy dynamics and thermodynamics of the SYK models.
 \begin{figure}[H]
\begin{minipage}{0.45\linewidth}
\centerline{\includegraphics[height=2.5cm]{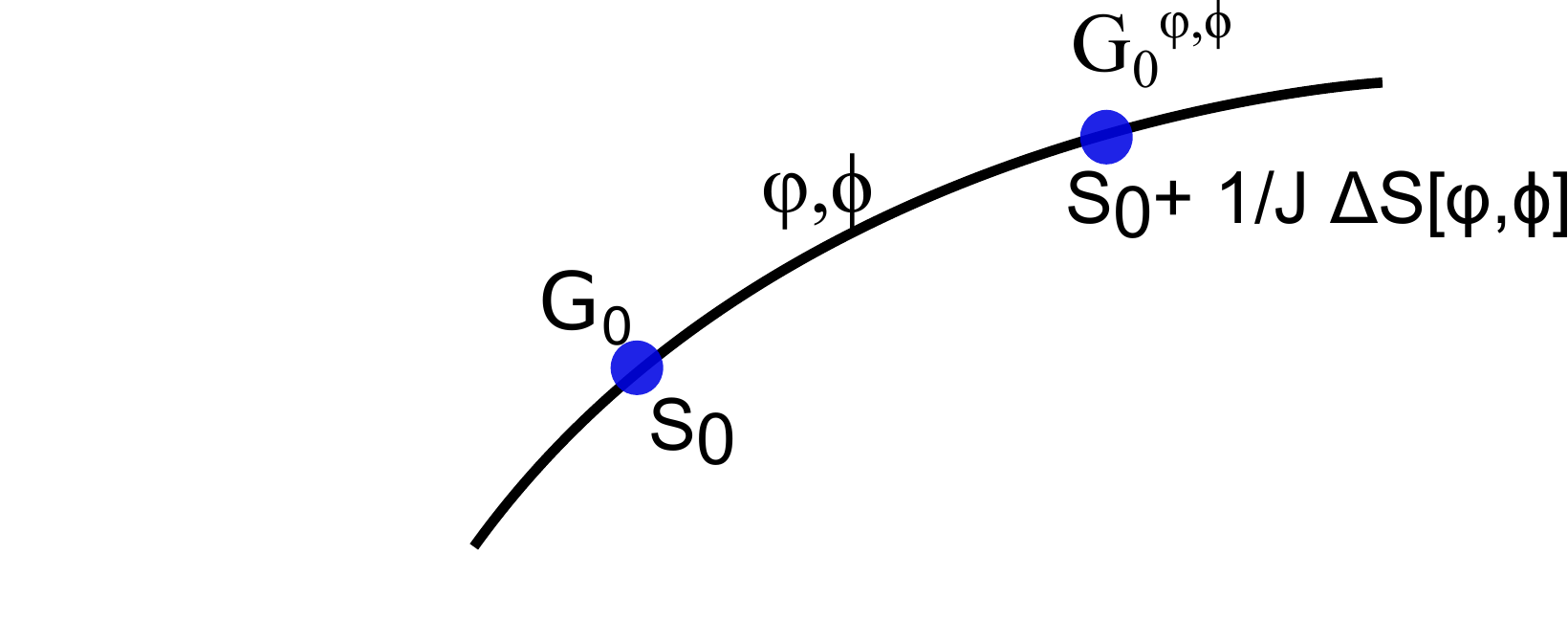}}
\end{minipage}
{\boldmath$\longleftrightarrow$}
\begin{minipage}{0.45\linewidth}
\centerline{\includegraphics[height=2.5cm]{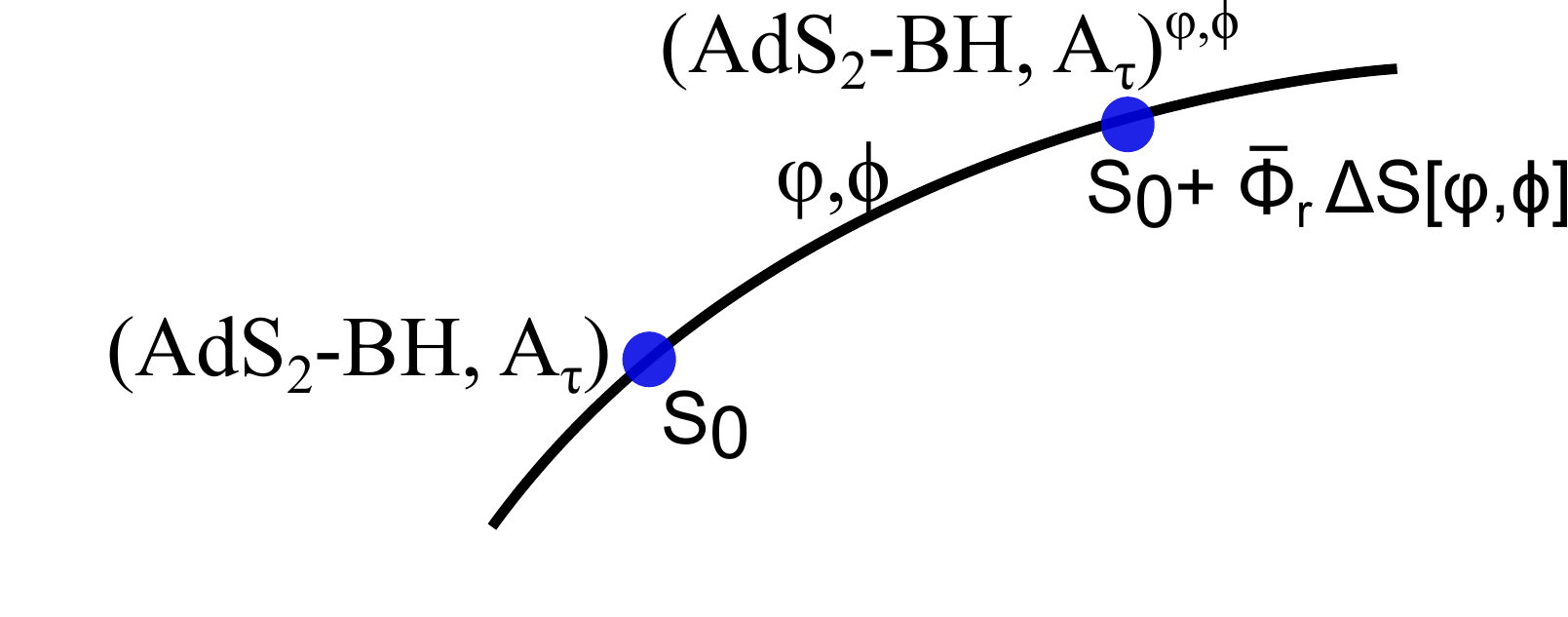}}
\end{minipage}
\caption{\footnotesize {The left panel represents the symmetry orbit of the classical large $N$ solution for the fermion bilocal $G_0(\tau_1, \tau_2)$, in the charged SYK model at the IR fixed point $J= \infty$. The symmetry group, {$\cal G$} is defined as the group of transformations generated by {\it Diff} and local U(1) transformations. The classical solution is invariant under ${\cal H}$= $SL(2,R) \times U(1)_{global}$ $\in$ {$\cal G$}. The orbit represents the Nambu Goldstone modes belonging to the coset ${\cal G}/{\cal H}$.  The right panel represents the symmetry orbit of \ads black hole under large diffeomorphisms. Points of the orbit parameterize the same coset as on the left, and are represented by the asymptotically AdS$_2$ spacetimes with horizon \eq{aads2intro} and $A_\tau$ given by \eq{eq:generalAt}}}
\label{fig-diff-gauge}
\end{figure}

In this paper, we explore possible origins of the near-\ads geometries in three dimensions. The original motivation came from the observation that the asymptotically \ads geometries \eq{aads2} of \cite{Mandal:2017thl} can be obtained as an induced metric on a domain wall from the asymptotically AdS$_3$ geometries of Brown and Henneaux \cite{brown1986, Banados:1998gg}.  In this paper, we find that a more useful approach appears to be Kaluza-Klein (KK) reduction from 3D Einstein gravity in a black hole phase, which leads to the JT-type 2D bulk model \cite{Almheiri:2014cka, Maldacena:2016upp}. A vanishing limit of the KK radius in our case corresponds to an exact \ads symmetry (although such a limit is singular), while a small non-zero KK radius denotes a near-\ads configuration. Under the KK reduction, the KK radius gets identified with the dilaton of the JT model. The breaking of the \ads symmetry by a non-zero dilaton therefore gets a geometric understanding in the KK scenario.

 The main new ingredient in our model is the treatment of the bulk dual of the charged SYK model. We find that in the 3D picture, incorporating the $U(1)$ charge comes naturally by including an abelian Chern-Simons term.  We add to 3D Einstein gravity a U(1) Chern-Simons term as well as a term which couples the two theories.  The KK reduction to 2D gives a BF theory coupled to the JT model with an appropriate coupling between the two, which reproduces the effective action of the pseudo-Goldstone modes of the charged SYK model. The 2D gauge fields are now acted on by large diffeomorphisms as well as large gauge transformations. The configuration space is represented by the right panel of Figure \ref{fig-diff-gauge}. In the near \ads background mentioned above, these configurations well capture the pseudo NG modes of the charged SYK model and reproduce the effective action \eq{total-s-eff-intro} (see the detailed summary below).

We should remark here that the bulk duals mentioned above do not purport to be a bulk dual of the full SYK model, rather they represent the `soft' sector \eq{soft}.  It is not clear if there {\it is} a local bulk dual for the full SYK model \cite{Gross:2016kjj,Gross:2017aos}. For proposals of a bulk dual for the massive modes of SYK model from a 3D viewpoint, see \cite{Das:2017pif, Das:2017hrt, Das:2017wae} \footnote{It would be interesting to explore possible connections   between this approach and the KK reduction studied in our present   work.}. Note that the original motivation for considering JT models in \cite{Almheiri:2014cka, Maldacena:2016upp} also comes from higher dimensions, namely from AdS$_d$ extremal black hole geometries ($d\ge4$) which have an \ads near-horizon geometry, see also \cite{STMV}. The breaking of the \ads isometry in that case is represented here by the physics away from the near-horizon region. As remarked above, in our case, of KK reduction from 3D, it is rather different; the breaking of the \ads symmetry is measured by a non-zero KK radius. 
\subsection*{Summary and organization of results}
In Section \ref{sec:charged-syk-review} we recapitulate the charged SYK model. We closely follow \cite{Davison:2016ngz} in this review.  

In Section \ref{sec:KK} we consider the KK reduction of Einstein-U(1) Chern Simons theory in 3D.  The salient points are:

\begin{itemize}

\item KK reduction of Einstein theory in 3D (with negative cosmological constant) reduces to a generalized Jackiw-Teitelboim model \cite{Jackiw:1984je, Teitelboim:1983ux}. The original references for this are \cite{Achucarro:1993fd, Hartman:2008dq, Castro:2008ms, Castro:2014ima}; see also the recent papers \cite{Papadimitriou, Mertens:2018fds, Gonzalez:2018enk} which use these ideas in contexts similar to that of the present paper. 
\item 
The dilaton comes from the radius of the KK direction; as emphasized above, a small non-zero KK radius corresponds to a small symmetry breaking parameter of \ads symmetry, giving a geometric meaning to such a role played by the dilaton in the JT model.
\item 
The KK reduction of the BTZ black hole is a 2D black hole with a dilaton solution. The 2D theory also admits the asymptotically \ads solutions \eq{aads2intro} which are obtained by applying large diffeomorphism to the 2D black holes\footnote{These are KK reductions of BTZ black holes with large diffeomporphisms in 3D in the presence of {\it non-normalizable deformations} (see appendix \ref{app:oxidation}) and hence, are  not a subset of Brown-Henneaux geometries.}. We discuss why the 3D AdS soliton is not relevant for our discussion. 
\item In Section \ref{KK-validity} we discuss the validity of the KK reduction. We show that the energy regime in which  KK reduction  is valid coincides with the IR regime of SYK. 
\item  In Section \ref{sec:charged-bulk} we introduce a U(1) Chern-Simons theory with a term that couples it to the Einstein gravity. We reduce the 3D Chern-Simons plus the coupling term to 2D. This KK reduction leads to a BF-type theory in 2D. The addition of the coupling between Chern-Simons field and gravity ensures a non zero field strength in 2D theory.  

\end{itemize}
Section \ref{sec:combo} contains the combined 2D action.  The salient features are,

\begin{itemize}
\item The bulk action for the holographic dual of low energy sector of charged SYK is presented in Section \ref{sec:combo}. We discuss the equation of motion and class of solutions in this section. 
\item 
The generalized JT model has contribution from the KK gauge fields. We show that a consistent truncation exists under which the KK gauge fields can be set to zero, after which the 2D theory becomes precisely the JT model. In the context of the BTZ solution, this truncation amounts to setting the angular momentum to zero.
\item{To account for thermal effects in gravity we take a black hole geometry as the reference geometry. This solution is the dimensional reduction of non-rotating BTZ spacetime in 3D gravity.} 
\end{itemize}

In the Section \ref{sec:soft-mode}, we present  derivation of boundary effective action from 2D gravity. We closely follow \cite{Maldacena:2016upp} to discuss the action of the soft modes from the bulk. The focus of the section is,
\begin{itemize}
\item{We discuss two points of view to look at the N\ads gravity. In this paper we adopt the first point of view (explained in section \ref{eader}). We discuss the physical origin of the action for the soft modes.}
\item{The soft mode action is eventually a boundary action and thus the boundary behavior of the fields matter the most in this computation. We, therefore, present a careful study of boundary fall off and boundary conditions on all the fields and accordingly propose covariant counter terms so as to keep the on shell action finite.}
\item Based on the boundary dynamics, in Section \ref{sec:schwarzian}, we obtain an effective action, which represents the action of pseudo Nambu-Goldstone modes of the charged SYK model. It  is given by a Schwarzian plus a sigma model term. The Schwarzian action is  already known from \cite{Maldacena:2016upp}. The sigma model in the context of charged SYK was proposed in \cite{Davison:2016ngz}.
\item The effective action we derive from the bulk agrees with the field theory effective action given in \cite{Davison:2016ngz}.  We conclude this section with appropriate identification of parameters in the bulk and boundary theory.
\end{itemize}

In section \ref{sec:chaos}, we compute chaotic correlators from the bulk. We reproduce the result that there is no contribution from ``boundary photons'' to the Liapunov exponent. 
\section{The charged SYK model\label{sec:charged-syk-review}}
The original SYK model involved Majorana fermions. In
\cite{Davison:2016ngz}, a generalized SYK model with Dirac fermions with a global $U(1)$ symmetry was introduced.  The model is given by a
hamiltonian
\begin{equation}
H = \sum_{i_1, ..., i_q}
  j_{i_1, ..., i_q} \psi^\dagger_{i_1}... \psi^{\dagger}_{i_{q/2}}
  \psi_{i_{q/2+1}}... \psi_{i_{q}}, \nonumber \\   
1\le i_1 <...< i_{q/2}\le N \quad \& \quad 1\le i_{\frac{q}{2}+1} <...< i_{q} \le N~,
\label{csyk_hamiltonian}
\end{equation}
where $q$ is an even number and $j_{i_1, ..., i_q}$ are complex Gaussian random variables with
\[
 j_{i_1, ..i_{q/2},i_{q/2 +1}.., i_q} = j^*_{i_{q/2 +1}.., i_q,i_1, ..i_{q/2}},
\quad \langle|j_{i_1..i_q}|^2\rangle=\frac{J^2((q/2)!)^2}{N^{q-1}}.
\]
\subsubsection*{Effective action and Schwinger-Dyson equations }
The Euclidean action for this model is
\begin{equation}
S = \int d\tau \left[ \frac{1}{2} \psi_i^{\dagger} (\partial_\tau - \mu)\psi_i\  - H \right]~,
\end{equation}
where $\mu$ denotes the chemical potential. The large $N$ limit is characterized by Schwinger-Dyson equations,
which can be derived from the following effective action in terms of
the bilocal field $\tilde G(\tau_1,\tau_2)=\frac{1}{N} \sum_i
\psi^{\dagger}_i(\tau_1)\psi_i(\tau_2)$ and an auxiliary field $\tilde
\Sigma(\tau_1,\tau_2)$
\begin{equation}
 S[\tilde G, \tilde \Sigma] = -\frac{N}{2}\ \text{Tr
   ln}\Bigg[\delta(\tau_1-\tau_2)\ \left(-\frac{\partial}{\partial
     \tau_2} + \mu\right) - \tilde
   \Sigma(\tau_1,\tau_2)\Bigg]\hspace{5cm}\nonumber
 \\ \hspace{3cm}-\frac{N}{2} \int d\tau_1 d\tau_2 \left[ \tilde
   \Sigma(\tau_1,\tau_2) \tilde G(\tau_2,\tau_1) + J^2(-1)^{q/2}
   \left(\tilde G(\tau_1,\tau_2)\right)^{q/2} \left(\tilde
   G(\tau_2,\tau_1)\right)^{q/2}\right]~.
 \label{csyk_effective_action}
\end{equation}
In fact, the original fermion integral with a quenched average over the disorder reduces, in the large $N$ limit, to a path integral over the variables   $\tilde G(\tau_1,\tau_2)$ and $\tilde \Sigma(\tau_1,\tau_2)$ with the above action \footnote{Our expression for the effective action differs slightly   from that of \cite{Davison:2016ngz} because of different conventions   for the bilocal fields and the conserved charge.}. The large $N$ limit of the above action is characterized by the following Schwinger-Dyson equations,
\begin{eqnarray}
 \Sigma(\tau) &=& - (-1)^{\frac{q}{2}} q
 J^2\ [G(\tau)]^{q/2}[G(-\tau)]^{q/2-1}\label{1scdy} \\ G(iw_n) &=& [iw_n   + \mu - \Sigma(i w_n)]^{-1}\label{2scdy}~,
\end{eqnarray}
here $w_n$ denotes the Matsubara frequency and $\tau = \tau_1-\tau_2$ (here $\Sigma$ and $G$ denote on-shell fields).
\subsection{Symmetry transformations}
In the IR regime defined by $w_n, \mu \ll J$ \cite{Davison:2016ngz} one can drop the $(iw_n + \mu)$ term from \eqref{2scdy}, leading to an emergent time reparameterization and U(1) gauge symmetry. 

The symmetry transformations of the charged SYK model, at strong coupling (the IR regime), consist of (a) diffeomorphisms parameterized by $f:\taut= f(\tau) \in$ {\it  Diff} \footnote{{\it Diff} represents ${\it Diff}\ S^1$ or ${\it Diff}\ R^1$,   depending on whether the system is at a finite temperature or zero   temperature.} as well as (b) gauge transformations parameterized by
$\exp[i\phi(\tau)]$ $\in U(1)_{\rm gauge}$. We will denote the combined
symmetry group as ${\cal G}$
\footnote{The action of {\it Diff} and $U(1)_{\rm gauge}$ do not
  commute. The parameterization of the combined symmetry
  transformation follows here the convention that the gauge
  transformation is performed first, followed by a {\it Diff}
  transformation, as in \eq{new-goldstone}.} which acts on the bilocal
meson variable $G(\tau_1, \tau_2) =(1/N)\ \psi^\dagger_i(\tau_1)
\psi_i(\tau_2)$ and $\Sigma(\tau_1,\tau_2)$ as follows:
\begin{align}
\tilde G(\taut_1, \taut_2) d\taut_1^\Delta d\taut_2^\Delta = G(\tau_1,
\tau_2)d\tau_1^\Delta d\tau_2^\Delta \exp[i\tilde\phi(\tilde\tau_2) - i
  \tilde\phi(\tilde\tau_1)]~,\nonumber\\
  \tilde \Sigma(\taut_1, \taut_2) d\taut_1^{(1-\Delta)} d\taut_2^{(1-\Delta)} = \Sigma(\tau_1,
\tau_2)d\tau_1^{(1-\Delta)} d\tau_2^{(1-\Delta)} \exp[i\tilde \phi(\tilde\tau_2) - i
  \tilde \phi(\tilde\tau_1)]~.
\label{new-goldstone}
\end{align}
Infinitesimally, for small $\phi(\tau)$ (here $\tilde \phi(\tilde \tau(\tau)) = \phi(\tau)$) and small $\ep(\tau)
\equiv f(\tau)- \tau$,
\begin{align}
\delta G &= \delta_\ep G + \delta_\phi G~,
\nonumber\\
\delta_\ep G(\tau_1, \tau_2) &=\left[ \ep(\tau_1)\del_{\tau_1}+ \Delta \ep'(\tau_1)
+ \ep(\tau_2)\del_{\tau_2}+ \Delta \ep'(\tau_2) \right]G(\tau_1, \tau_2)
\label{delta-ep}~,
\\
\delta_\phi G(\tau_1, \tau_2)&= i \left(\phi(\tau_2)- 
\phi(\tau_1)\right)G(\tau_1, \tau_2)~.
\label{delta-Lam}
\end{align}
The vacuum solution $G_0(\tau_1, \tau_2)$ at strong coupling
\footnote{For example, at zero temperature,
\[ G_0(\tau_1, \tau_2)= G_0(\tau) = \left\{
\begin{array}{l} - C |\tau|^{-2 \Delta},~~\tau>0 
\\  C e^{-2\pi {\cal E}}|\tau|^{-2 \Delta}, ~~\tau<0  \end{array}
\right.
\]
where ${\cal E}$ is the `spectral asymmetry paramater' satisfying ${\cal E}$ =lim$_{_{T\to 0}}$ $|\del \mu/\del T|_{_Q}$, $\tau= \tau_1-\tau_2$, $Q$ and $\mu$ are the U(1) charge and the corresponding chemical potential respectively and $C$ is a constant. For more details and for the finite temperature two-point function, see \cite{Davison:2016ngz}. \label{ftnt-2pt}} breaks the symmetry from ${\cal G}$ to ${\cal H}= \slr \times U(1)_{\rm  global}$. The Nambu-Goldstone (NG) modes are therefore parameterized by the coset space $\left(f(\tau), \phi(\tau) \right)$ $ \in {\cal  G}/{\cal H}$. As in the uncharged case, the strict limit $J= \infty$ is singular since the NG modes have zero action.  In order to make sense of the theory, one needs to turn on a small value of an irrelevant coupling $1/J$, thereby explicitly breaking the symmetry ${\cal G}$. The NG modes now become pseudo Nambu-Goldstone modes, with an action which is a generalization of the Schwarzian action of the uncharged model \cite{Davison:2016ngz}. This action is given by \eq{total-s-eff-intro}, \eq{schwarzian-intro}, \eq{sigma-model-intro}.
\section{Kaluza-Klein reduction of 3D gravity and Chern-Simons\label{sec:KK}}
The bulk dual to the SYK model with Majorana fermions have been previously discussed in \cite{Maldacena:2016upp, Mandal:2017thl}. We want to investigate the possible holographic dual to charged SYK using a Kaluza-Klein (KK) reduction of 3D gravity. Below, we propose the bulk dual to be a KK reduced action derived from 3D Einstein gravity with negative cosmological constant and a U(1) Chern-Simons with a source term. To this end, we perform KK reduction of one of the coordinates on a circle $S^{1}$. It is convenient to look at the reduction of AdS$_3$ and Chern-Simons sector separately.
\subsection{3D Einstein gravity\label{sec:KK-grav}}
We begin with the three dimensional Einstein action with a negative cosmological constant,
\begin{equation}\label{S3Dgrav}
 S_{\rm{3DGrav}} = -\frac{1}{16 \pi G_3} \int_{M_3} d\tu dz dy \ \sqrt{g^{(3)}} \left(R^{(3)} + \frac{2}{l^2}\right) - \frac{1}{8 \pi G_3} \int_{\del M_3} d\tu dy \ \sqrt{h}\ \left(\mathcal{K}^{(3)}-\f 1 l\right)~.
\end{equation}
The 3D manifold $M_3$ is parameterized by $x^M = \{\tau, y,z\}$, with a boundary at $z= 0$,
which we will regulate by a cut-off $z=\delta$. The coordinates on the  boundary $\del M_3$ are given by $x^\a= \{\tau, y\}$ with the induced metric $h_{\alpha\beta}$. We have taken the cosmological constant to be $\Lambda = - \f1{l^2}$, where $l$ denotes AdS length scale. The boundary term above are the usual Gibbons-Hawking contribution and a counter term to have finite boundary current. We will work with Euclidean signature throughout this paper. 

Varying the above action with Dirichlet boundary condition on the metric yields Einstein equations,
\begin{align}
R^{(3)}_{MN}- \f12 \left(R^{(3)}+  \f{2}{l^2}\right)g^{(3)}_{MN}=0~.
\label{einstein-eq-3D}
\end{align}
\subsubsection{Kaluza-Klein reduction to 2D}
To perform a KK reduction of the 3D gravity we begin with the ansatz,
\begin{align}
ds_3^2 &= g_{\mu\nu} dx^\mu dx^\nu + l^2 \Phi^2(d\theta + B_{\mu}  dx^\mu)^2
~,~ dy=l d\theta~,~  \theta\equiv \theta + 2\pi~,
\label{KK-reduction}
\end{align}
where all the fields in \eqref{KK-reduction} are functions of $x^\mu=\{\tu, z\}$ which parameterize the 2D manifold $M_2$.  The metric on $M_2$ is represented by $g_{\mu\nu}$, $B_\mu$ is a vector field  and $\Phi$ denotes scalar in $M_2$. Using \eq{KK-reduction}, we can derive the following equations
\begin{eqnarray}
 R^{(3)} &=& R^{(2)} - 2\ \frac{1}{\Phi} \nabla^2 \Phi- \frac{l^2}{4} \Phi^2 \tilde F^2~,~\mathcal{K}^{(3)} = \mathcal K^{(2)}+ n^\mu \frac{1}{\Phi}D_\mu \Phi~,~\nonumber\\
\sqrt{g^{(3)}} &=&l\ \Phi \sqrt{g^{(2)}}~,~ \sqrt{h} =l\ \Phi \sqrt{\gamma}~.
\label{3d-2drelations}
\end{eqnarray}
Here $\tilde F^2= \tilde F_{\mu \nu }\tilde F^{\mu \nu}$ is the field strength for KK gauge field $B_\mu$ such that $\tilde F_{\mu\nu} = \del_\mu B_\nu - \del_\nu B_\mu$. The determinant of the induced metric on the 1D boundary $\del M_2$ is $\gamma$ and $n^{\alpha}$ is a unit normal to the boundary. Using these, the 3D action \eq{S3Dgrav} reduces to a  Jackiw-Teitelboim (JT) \cite{Jackiw:1984je, Teitelboim:1983ux} action plus a term due to KK field strength: 
\begin{equation}\label{eq:s2dgrav}
S_{\rm{2DGrav}} = -\frac{1}{16 \pi G_2} \int_{M_2} d\tu dz\ \sqrt{g}\ \Phi \left(R + \frac{2}{l^2} - \frac{l^2}{4} \Phi^2 \tilde F^2\right) - \frac{1}{8 \pi G_2} \int_{\del M_2} d\tu \sqrt{\gamma}\ \Phi\ \left(\mathcal K-\f 1 l\right) ~.
\end{equation}
We have dropped the superscripts for now  as we will only be working in $2$ dimensions, we will bring them back whenever necessary. Note that $G_3= 2\pi l G_2$. The details of the above reduction can be found in Appendix \ref{appen:KK}.
\subsubsection{KK reduction of specific solutions}
\subsubsection*{BTZ black holes} The equation of motion of 2D gravity can either be derived from the variational principle in 2D action or they can equivalently be derived from the equation \eq{einstein-eq-3D} and \eq{KK-reduction}. We discuss the 2D equations and their solution in detail in section \ref{sec:combo}. However, we already know the BTZ black hole \cite{Banados:1992wn} as the solution to \eq{einstein-eq-3D}, where the metric is given as,
\begin{align}
ds_{\rm{BTZ}}^2 = \left(\f {l^2}{z^2}-M l+\frac{J^2 z^2}{4 l^4}\right)d\tu^2 + \left(\f {l^2}{z^2}-M l+\frac{J^2 z^2}{4 l^4}\right)^{-1}\frac{l^4 }{z^4}dz^2 + \frac{l^2 a^2}{z^2} \left( d\theta-i\frac{J}{2} \frac{z^2}{l^4} d\tau\right)^2~,
\label{rotating-btz}
\end{align}
here, $z \in (0, \infty)$ and the coordinates $\tau= [0, \b]$, and
$\theta =[0, 2\pi]$ represent the conformal boundary $S^1 \times S^1$. The mass $M$ and angular momentum $J$ are the black hole parameters. Note that $2\pi a/\beta$, by definition, represents the ratio of the asymptotic proper radius of $\theta$ and $\tau$ circles. 

We can readily identify the solution to the 2D equations, for the action \eq{eq:s2dgrav}, from the above BTZ spacetime. They are, 
\begin{eqnarray}
 && ds^2_{2D} = \left(\f {l^2}{z^2}-M l+\frac{J^2 z^2}{4 l^4}\right)d\tu^2 + \left(\f {l^2}{z^2}-M l+\frac{J^2 z^2}{4 l^4}\right)^{-1}\frac{l^2 a^2}{z^4}dz^2~,\nonumber\\
 &&  \Phi =\frac{ a}{z}~.\nonumber\\
 && B_\tau =-i\frac{J z^2}{2 l^4} ~, ~B_z=0~.\label{2dschwar}
\end{eqnarray}
 Note that, only when  the 3D fields are independent of the compactified direction $y$, the mapping of solutions under KK reduction is possible. This means that the 3D solutions of the form \eq{KK-reduction}, where the metric components depend on the coordinates $\{\tau, z\}$ only, provide a solution to the 2D equation of motion directly.

 The non-rotating BTZ black hole in the  Fefferman-Graham gauge is given as, 
\begin{eqnarray}
 ds_{\rm{3DBTZ}}^2 = \frac{l^2}{4 z^2}\left(\frac{M z^2}{l}-1\right)^2 d\tu^2 +\frac{l^2}{z^2}dz^2+\frac{l^2 a^2}{4 z^2}\left(\frac{M z^2}{l}+1\right)^2d\theta^2~\label{eq:btz-nonrotating}
\end{eqnarray}  
with $\tau\in[0,\beta]$, $\theta \in [0, 2\pi ]$ and $\beta\sim 1/\sqrt{M}$.

From this 3D metric, we can write the solutions in 2D gravity to be, 
\begin{eqnarray}
&& ds_{\rm{2D}}^2 = \frac{l^2}{4 z^2}\left(\frac{M z^2}{l}-1\right)^2 d\tu^2 +\frac{l^2}{z^2}dz^2, 
\nonumber\\
 &&  \Phi =\frac{a}{2z}\left(\frac{M z^2}{l}+1\right)~.\label{eq:2d-btz-nonrotating}
\end{eqnarray}
The above solution represents AdS$_2$ black hole with dilaton $\Phi$. 
\subsubsection*{3D AdS-soliton}
The equations \eq{einstein-eq-3D} (and hence equations of motion for
the action \eq{eq:s2dgrav}) admit another well-known solution, the
AdS soliton \cite{Horowitz:1998ha}, \begin{eqnarray}
  ds_{\rm{soliton}}^2 = \frac{l^2}{4 z^2}\left(\frac{M
    z^2}{l}+1\right)^2 d\tu^2 +\frac{l^2}{z^2}dz^2+\frac{ l^2a^2}{4
    z^2}\left(\frac{M
    z^2}{l}-1\right)^2d\theta^2~\label{eq:3D-soliton}
\end{eqnarray}
with $\tau\in[0,\beta]$, $\theta \in [0, 2\pi ]$ and $a\sim 1/\sqrt{M}$. The radial  coordinate is $z\in (0, \infty)$ and the other two coordinates have the period as, $y\in(0,2\pi l)$ and $\tau\in (0,\beta)$. It is easy to show that for low temperatures the free energy of the soliton solution is lesser compared to that of BTZ and thus the soliton phase is preferred for low temperatures.

Let us now look at the 2D solutions derived from 3D solutions of the type \eq{eq:3D-soliton}. The solution from the BTZ is given in equation \eq{eq:2d-btz-nonrotating} and the solution from  soliton will be,
\begin{align}
\label{eq:2d-soliton}
ds_{\rm{2D}}^2 = \frac{l^2}{4 z^2}\left(\frac{M z^2}{l}+1\right)^2 d\tu^2 +\frac{l^2}{z^2}dz^2~, 
~  \Phi =\frac{a}{2z}\left(\frac{M z^2}{l}-1\right)~.\end{align}
\subsubsection*{An apparent puzzle}
When one considers KK reduction along the $\theta$-circle, then the KK radius, measured by the proper radius of the $\theta$-circle at the boundary, must be smaller than various length scales of relevance of physics in the lower dimension. In particular, this proper radius must be smaller than that of the $\tau$-circle, which measures the inverse temperature. Now, it is well-known that under this circumstance, the thermodynamically favourable solution is the AdS soliton \cite{Horowitz:1998ha} which has a contractible $\theta$-circle and not the BTZ black hole which has a contractible $\tau$-circle. In that case, to arrive at the 2D black hole \eq{eq:btz-nonrotating} by KK reduction from 3D, one {\it has to start from the sub-dominant saddle point of   the Euclidean path integral}, namely the BTZ solution. At large $N$, there is a sense in which one can explore the perturbative neighbourhood of a sub-dominant saddle point since the tunneling amplitude to the dominant saddle point (the AdS soliton) is of order $e^{-O(N)}$. In doing so we are aware that there is a loss of unitarity up to non-perturbative effects of order $e^{-O(N)}$. There is a complementary argument which says that KK reduction from the 3D soliton solution along the $\theta$-circle does not in any case make sense. The reason is that for the AdS soliton, the KK circle contracts in the bulk; this leads to a vanishing dilaton in 2D (see \eq{eq:2d-soliton}) and consequently uncontrolled quantum fluctuations.

In this paper, we will take the viewpoint that we will consider the classical saddle point solution \eq{eq:btz-nonrotating} and its 2D reduction \eq{eq:2d-btz-nonrotating} and explore the geometries obtained by considering different boundary curves (we will explain later in section \ref{eader} how this is the same as orbits of this solution under large diffeomorphism), which represents the correct physics of the SYK model. The relevance of the AdS soliton is unclear from the point of view of the SYK model.
\subsubsection{Validity of the Kaluza Klein reduction}
\label{KK-validity}
Note that in \eq{eq:btz-nonrotating} if we take $z \to 0$ limit naively then $g_{\theta \theta}$ diverges which would appear to invalidate the KK reduction. However a sensible KK reduction can be obtained if we work with a finite cutoff $z=\delta$, in that case $g_{\theta\theta} \sim \left(\f{l}{\delta} a\right)^2$ up to numerical factors. Hence the
KK radius  is $R_{KK} \sim \f{l}{\delta} a$. For the KK reduction to be
valid, the typical energy $E$ of particles in a bulk correlator
should satisfy the relation
\[
E \ll  \f1{R_{KK}} \sim \f{\delta}{l} \f1a
\]
Note that the energy $E$ measured in the bulk is related to the energy
$\om$ in the boundary theory by the standard AdS/CFT equation: $E=
\f\delta{l} \om$. \footnote{This is related to the fact that the AdS
  metric induced on the boundary $z= \delta$ is $ds^2|_{_\delta}=
  \f{l^2}{\delta^2} \left(dx^\mu dx_\mu\right)$.} 

Later on in this note we will identify the coefficient $a$ (which is a
non-normalizable deformation of the dilaton) with $1/J$ (see
\eq{parameters}).  Hence the above condition for the
validity of the KK reduction reduces to
\[
\om \ll  J
\]
which is the same as the condition (see \eq{soft}) that we are
working in the IR regime of the SYK theory!
\subsection{3D Chern-Simons}
\label{sec:charged-bulk} 
In order to obtain a dual to charged SYK we must include gauge fields
in 3 dimensions besides gravity. The most natural candidate in the long wavelength limit   is a U(1)
Chern-Simons (CS) action in an asymptotically AdS$_3$ spacetime
(recall that the Maxwell term is irrelevant in the IR). The action
must have a boundary term for a well-defined variational
principle. The CS action, with appropriate boundary conditions and
boundary terms, following \cite{Kraus:2006wn}, is given by
\begin{align}
 S_{CS} &= \frac{i~k}{8\pi} \int_{AdS_3} AdA - \frac{k}{16 \pi}\ \int_{\d AdS_3} d\tu\ dy\ 
\sqrt{h}h^{\a\b}\ A_\a A_\b ~.
\label{kraus-action}
\end{align}
Here, $k$ is a real positive parameter. Although the boundary term introduces a coupling of the boundary $U(1)$ gauge field to 3D gravity, the $U(1)$ gauge field in the bulk is not coupled to gravity in the above action, and hence the field strength vanishes as usual. This forces the Wilson loop around the time circle to vanish in geometries where this circle is contractible.  As we will see below, this implies a vanishing chemical potential. Hence, in order to have the correct bulk dual to a theory with a non-zero chemical potential, we must introduce a coupling between the bulk metric and the bulk CS $U(1)$ gauge field. The simplest such coupling is 
\begin{align}
S_{Coupling}= -\f {i k}{4\pi}
\int_{AdS_3}{d\tau dz dy\sqrt{g^{(3)}}~ A_M J^M}  
\label{coupling}
\end{align}
where the $J^M$ is a given external source, which we will specify in detail shortly (see \eq{j-y}).

Next, we do a KK reduction along the $y$ direction, as before.
\subsubsection{KK reduction to 2D}
Under KK compactification along the $y$ direction, the gauge field is taken
to be of the form
\begin{equation}
A_M dx^M= A_\mu dx^\mu + l\ \rchi\ d\theta~.
\end{equation}
where $A_\mu$ and $\rchi$ denote vector and scalar fields in 2 dimensions. 

Unlike the dynamical gauge field, the external current $J^M$ 
can only be a scalar constant due to diffeomorphism invariance,
that is, 
\begin{align}
J^\tu=  J^z =0, \;  J^y= J_0  = \hbox{constant.}
\label{j-y}
\end{align}
Then from equation \eq{KK-reduction} for the metric, the  KK reduction of the action \eq{kraus-action} is,
\begin{align}
S_{\rm{2DGauge}}&=
\f{i k l}{2} \int_{M_2} \rchi F_{\tu z}d\tu dz-  J_0 \f {i k l}{2}\int_{M_2} \sqrt{g}~\Phi\rchi ~d\tu dz + \frac{i k l}{4}\int_{\del M_2} \rchi A_\tau d\tau   
\nonumber\\
& -\f{k l}{8} \int_{\del M_2} d\tau ~ \sqrt{\gamma}\ \Phi\ \gamma^{\tau \tau}\left( A_\tau^2+ \gamma_{\tau \tau}\left(\frac{\rchi}{\Phi}\right)^2 + \left(l\rchi B_\tau\right)^2 -2 l\rchi B_{\tau}A_\tau\right)~,
\label{eq:s2dgauge}
\end{align}
here $M_2$ denotes 2d bulk manifold and $\del M_2$ is the boundary given by $z=\delta$. We have used the definition $F_{\al\be} = \del_\al A_\be - \del_\be A_\al$.  
KK Reduction of the gravity sector as described in section \ref{sec:KK-grav} remains unaffected by the coupling \eq{coupling} to Chern-Simons fields.
\section{Combined action in 2D: bulk dual to charged SYK\label{sec:combo}}
As indicated in the introduction, a conserved $U(1)$ current in the boundary theory implies, by usual AdS/CFT correspondence, that the bulk theory should have a $U(1)$ gauge field. We propose the Kaluza-klein reduced CS and the JT action plus the coupling term \eq{coupling} discussed above as the bulk dual for the low energy sector of the charged SYK model. The reduced two dimensional action consists of two terms, 
\begin{equation}\label{S2D}
S_{\rm{2D}}=S_{\rm{2DGrav}}+S_{\rm{2DGauge}}
\end{equation}
where $S_{\rm{2DGrav}}$ and $S_{\rm{2DGauge}}$ are given by equations \eq{eq:s2dgrav} and \eq{eq:s2dgauge} respectively. 
\begin{align}
 S_{\rm{2D}}=& -\frac{1}{16 \pi G_2} \int_{M_2} d\tu dz\ \sqrt{g}\ \Phi \left(R + \frac{2}{l^2} -\frac{l^2}{4} \Phi^2 \tilde{F}^2\right) -\f{i kl}{2} \int_{M_2} d\tu dz\ \s{g}\ \rchi \left( J_0\ \Phi -F\right)   
 ~\nonumber  \\
& + \frac{i kl}{4}\int_{\del M_2} d u\ \rchi A_u -\f{k l}{8} \int_{\del M_2} d u \sqrt{\gamma}\ \Phi\  \gamma^{u u}\left( A_u^2+ \gamma_{u u}\left(\frac{\rchi}{ \Phi}\right)^2 + \left(l\rchi B_u\right)^2 -2l \rchi B_{u}A_u\right)\nn \\ 
&- \frac{1}{8 \pi G_2} \int_{\del M_2} d u \sqrt{\gamma}\ \Phi\ \left(\mathcal K-\f 1 l\right)
\label{full-action-2d}
\end{align}
In the above expression, $F=\f{1}{2}\ {\epsilon^{\mu\nu}} F_{\mu \nu}/{\s{g}}$ is a scalar function with $\epsilon^{\mu\nu}$ being the Levi-Civita symbol ($\ep^{01} = -\ep^{10} = 1$). The boundary manifold $\del M_2$ is assumed to be given by $(\tau,z) = (\tau(u),z(u))$ where $u$ is the coordinate of the 1-dimensional boundary manifold and $A_u$, $B_u$ and $\gamma^{uu}$ are induced gauge fields and metric on this manifold. Hear after we drop the subscript $M_2$ and $\del M_2$ from the integrals and only work with the 2D theory.

\subsection{Equations of motion and solutions}
\label{sec:all-equations}
The classical equations of motion from the above action are, 
\begin{subequations}
\begin{align}\label{phivar}
&R + \frac{2}{l^2} - \frac{3 l^2}{4}\ \Phi^2\ \tilde F^2 -8 \pi G_2 i kl~ \rchi J_0= 0~,
\\
\label{Bmuvar}
& \nabla_\mu (\sqrt{g}\ \Phi^3 \tilde F^{\mu \nu}) = 0~,
\\
\label{gvar}
&\left(\nabla_\mu \nabla_\nu \Phi - g_{\mu\nu} \nabla^2 \Phi + \frac{g_{\mu\nu}}{2}\left(\f{2}{l^2}-8\pi G_2 i k \rchi J_0\right) \Phi\right) + \frac{\Phi^3 l^2}{4} \left(-\frac{1}{2}g_{\mu\nu} \tilde F^2 +2 g^{\alpha \beta} \tilde F_{\alpha \mu} \tilde F_{\beta \nu}\right) = 0 ~,\\
&\partial_\tau\rchi=0~, \quad \partial_z \rchi=0~\label{eq:chieq}, \\
& F=  J_0~\Phi\label{eq:sourceeq}. 
\end{align}
\end{subequations}
Note that it is consistent, in equations \eq{Bmuvar} and \eq{eq:chieq}, to put
\begin{align}
\label{cons-trun}
& B_\mu=0~\text{and}~\rchi=0~
\end{align}
since all source terms for these fields involve themselves \footnote{It is important to mention here that any large gauge transformations of the field $B_\mu$ will not affect our results due to the consistent truncation $\rchi=0$.}(note that $\tilde F_{\mu \nu} = \del_\mu B_\nu - \del_\nu B_\mu$). With this,
the equations reduce to,
\begin{subequations}
\begin{align}
&R + \frac{2}{l^2} = 0,\label{rp2}\\  
&\nabla_\mu \nabla_\nu \Phi - g_{\mu\nu} \nabla^2 \Phi + \frac{g_{\mu\nu}}{l^2} \Phi = 0 \label{dileom},\\  
&F= J_0~\Phi \label{eq:trunc}.  
\end{align}
\end{subequations}
A solution to these equations is the Poincare \ads metric, 
\begin{equation}\label{ads}
 ds^2_{\rm ads} = \frac{l^2}{\tilde z^2} \left( d\tilde z^2 + d\tilde \tau^2 \right) 
\end{equation}
in which, using Eq. \eqref{dileom}, the dilaton takes the following form,  
\begin{equation}
\label{eq:generalPhi}
\tilde \Phi(\tilde \tau, \tilde z) = \f{a + b\ \tilde \tau + c\ (\tilde \tau^2 +\tilde z^2)}{\tilde z}
\end{equation}
Here $a$, $b$ and $c$ are integration constants. We work in the gauge $\tilde A_{\tilde z}(\tilde \tau, \tilde z)=0$ and with this choice, Eq. \eqref{eq:trunc} leads to, 
\begin{equation}
 \tilde A_{\tilde \tau}(\tilde \tau, \tilde z) = J_0 l^2 \left(\frac{a + b \tilde \tau + c \tilde \tau^2}{2 \tilde z^2} + c \log(\tilde z)\right) 
\end{equation}
It is easy to show that a coordinate transformation of the form , 
\begin{align}\label{eq:exact-large-diff}
\tilde\tau(\tau, z) = f(\tau) - \frac{2 z^2 f''(\tau) f'(\tau)^2}{4 f'(\tau)^2 + 
z^2 f''(\tau)^2}, \quad \tilde z(\tau, z) = \frac{4 z f'(\tau)^3}{4 f'(\tau)^2 + 
z^2 f''(\tau)^2}~,
\end{align}
give rise to asymptotically \ads solutions of Eqs. \eqref{rp2} \cite{Mandal:2017thl}. These are family of solutions parametrized in terms of the function $f(\tau)$ with the metric given as, 
\begin{equation}\label{aads2}
 ds^2_{\rm AAdS_2} = \frac{l^2}{z^2} \left( dz^2 + d\tau^2 \left( 1 - \frac{z^2}{2} \qty{f(\tau),\tau}\right)^2\right).
\end{equation}
As it is obvious, under \eqref{eq:exact-large-diff} the dilaton transforms as $\Phi(\tau, z)=\tilde \Phi(\tilde \tau, \tilde z)$ and the gauge fields, under the choice $\tilde A_{\tilde z}(\tilde \tau, \tilde z)=0$, transform as, 
\begin{eqnarray}
 A_\tau(\tau,z) &=& \tilde A_{\tilde \tau}(\tilde \tau, \tilde z) \f{\del \tilde \tau(\tau,z)}{\del \tau}\label{eq:generalAt}\\
 A_z(\tau,z) &=& \tilde A_{\tilde \tau}(\tilde \tau, \tilde z) \f{\del \tilde \tau(\tau,z)}{\del z}
\end{eqnarray}
Note that this diffeomorphism breaks the $A_z =0$ gauge condition, but we can do a simultaneous gauge transformation with the diffeomorphism that will restore this gauge condition.

\subsubsection*{A note on classification of large diffeomorphisms}
Large diffeomorphisms given in \eq{eq:exact-large-diff} are not bijective maps everywhere inside the bulk but only near the boundary hence they exactly map \ads boundary into boundary of asymptotically \ads. This map is good for our purpose as we will only need the near boundary behaviour of fields in this work. They are also in one to one correspondence with boundary time reparameterization functions $f(\tau)$ of the ${\it Diff}\ \mathbb{R}^1$ group. We would like to point out that the 2D black hole metric can be generated by taking $f(\tau)= (\beta/\pi)\ \tan(\pi \tau /\beta)$. Coordinate transformations changing the topology from \ads to black hole seems surprising. However, on a careful look we find that this is possible because these maps become singular at the horizon. Keeping this in mind we make a distinction between geometries produced by large diffeomorphism of \ads and large diffeomorphisms of 2D black hole. As we are interested in finite temperature systems we will always consider 2D black hole to be our reference geometry and work 
with this class of geometries parameterized by $f(\tau) = (\beta/\pi) \tan(\pi \varphi(\tau)/\beta)$ where $\varphi(\tau)$ belongs to ${\it Diff}\ \mathbb{S}^1$ group.

\subsubsection{2D black hole solutions with dilaton and gauge field}

We can check that if we plug in $f(\tau) = (\beta/\pi) ~ \tan(\pi \tau/\beta)$ in \eq{aads2} we obtain a  black hole solution to the Eq. \eq{rp2}, with its horizon at $z = \frac{\beta}{\pi}$,
\begin{equation}\label{2dbtz}
 ds^2_{\rm BH} = \frac{l^2}{z^2} \left( dz^2 + d\tau^2 \left( 1 - z^2\ \f{\pi^2}{\beta^2}\right)^2\right),
\end{equation}
and the dilaton in this background is
\begin{equation}\label{btzgendil1}
 \Phi(\tau, z)= \frac{(a \pi^2 + c \beta^2 )\ \left(\beta ^2+\pi ^2 z^2\right)}{2 \pi^2  \beta^2  z } + \left((a \pi^2 - c \beta^2 ) \cos \left(\frac{2 \pi  \tau }{\beta }\right) + b \pi \beta  \sin \left(\frac{2 \pi  \tau }{\beta }\right)\right)~\frac{\beta^2 -\pi^2 z^2 }{2 \pi^2  \beta^2  z }~,
\end{equation}
with the gauge field $A_\tau(\tau,z)$ in the gauge $A_z(\tau,z)=0$ is,
\begin{equation}\label{atgen1}
 A_\tau(\tau,z) = a_1(\tau)\ \f{1}{z^2} + a_2(\tau)\ \log(\frac{\pi z}{\beta}) + a_3(\tau) ~z^2 + a_4~.
\end{equation}
Here $a_i$'s are time dependent functions parameterized by $a$, $b$, $c$ and the source $J_0$ (note that these are the same constants as presented in Eq. \eq{eq:generalPhi}). Detailed expressions for these coefficients are presented in Appendix \ref{gena}. At this point we pause for a couple of comments,
\begin{itemize}
 \item We have imposed $ A_z(\tau ,z) = 0 $ gauge, this will restrict large gauge transformations to be functions of time only.
 \item $A_\tau(\tau,z)$ should vanish at the horizon of the black hole (which is required by Stokes' theorem for non-singular field strengths included at the horizon). This is  arranged by choosing the integration constant appropriately.
 \item The solution \eq{atgen1} has a log($\frac{z\pi}{\beta}$) term, this mode does not have an obvious AdS/CFT interpretation and gives divergences in the on-shell action. Thus we will switch off this mode by making a choice of parameters: $ c = a\pi^2/\beta^2,\ b=0\  (\implies a_2=0 )$
 \item Once the logarithmic mode is switched off ($a_2=0$), the leading order divergence is given by $a_1$, we will show that if we fix a Dirichlet boundary condition for the dilaton field on the boundary then $a_1$ is also completely fixed by the same boundary condition due to \eq{eq:trunc}.
 \item The finite contribution form the gauge field on the boundary will be identified with the chemical potential of the boundary theory up to multiplicative factors.
\end{itemize}

As mentioned above, we choose $ c = a\pi^2/\beta^2,\ b=0 $. This choice does two things for us it removes the $\log$ mode from the gauge field and it also makes the solutions time independent,
\begin{equation}\label{btzgendil}
 \Phi(\tau, z) = \frac{a}{z} + \frac{a \pi^2}{\beta^2}\ z 
\end{equation}
and
\begin{equation}\label{atgen}
 A_\tau(\tau,z) = \left(\frac{a  J_0 l^2}{2 }\right)\ \f{1}{z^2} + \left(\frac{a J_0 l^2\pi^4}{2 \beta^4}\right) z^2 - \frac{aJ_0 l^2 \pi^2}{ \beta^2}.
\end{equation}

This completes our discussion about the bulk action and various classical solutions to the Eqns., \eq{rp2}-\eq{eq:trunc}.  In the next section we derive the action for the soft modes in the bulk.
\section{Effective action for pseudo Nambu-Goldstone modes}\label{sec:soft-mode}
In this section, we recall the derivation of effective action for pseudo Nambu Goldstone (pNG) modes in general theories.

Consider a scenario where we have an action $S_0(G)$ symmetric under a group $\mathcal{F}$ whose elements are denoted by $f$. The classical vacua can be parameterized as $G_0^f$ where $G_0$ is the reference classical vacuum and $f$ is a group element, as $S_0$ is symmetric under $\mathcal{F}$, $S_0 (G_0) = S_0 (G_0^f)$. Now if we explicitly add a small symmetry breaking term $\Delta S$ such that total action $S_T = S_0 + \Delta S$ is not symmetric under $\mathcal{F}$ ($S_0(G_0) + \Delta S(G_0) \neq S_0(G_0^f) + \Delta S(G_0^f)$), then $G_0^f$ are not exact zero modes of $S_T$ and they represent pseudo Nambu Goldstone (pNG) modes. The action for these pNG modes can be computed by taking a difference between a reference vacuum $G_0$ and all other vacua $G_0^f$ of $S_0$
\begin{equation}
S_{pNG}(f) = [S_0(G_0^f) + \Delta S(G_0^f)] - [S_0(G_0) + \Delta S(G_0)]~,\nonumber \\
S_{pNG}(f) = \Delta S(G_0^f) - \Delta S(G_0), ~\hspace{3.5cm}
\label{SpNG}
\end{equation}
$\Delta S(G_0)$ does not contain any dynamical field. Note that $G_0^f$ is a solution of $S_0$ and in general is an off shell configuration for $S_T$.

It is also possible that there exists a subgroup $\mathcal{H}$ of $\mathcal{F}$ whose elements are denoted by $h$, for which $G_0^h = G_0$, which will mean $\Delta S(G_0) =  \Delta S(G_0^h)$. This will indicate $S_{pNG}$ is symmetric under $\mathcal{H}$.

Let us take a  look at how this works out in the SYK model. At the conformal fixed point, ${\it Diff}$ group which is represented by $f(\tau)$ is an exact zero mode of the SYK action (it is easy to see this when SYK action is written in terms of bilocals $\Sigma$ and $G$ \cite{Maldacena:2016hyu}); however at this point the theory is singular. In order to get a well defined theory we have to explicitly break this symmetry by bringing back the term proportional to $1/J$ this leads to an action for pNG modes exactly as given in the above discussion. For the conformal two point function of SYK, $G^h = G$ when $h$ is element of $SL (2,\mathbb{R})$. Thus from above discussion we expect that the action for pNG modes would have global $SL(2,\mathbb{R}) $ symmetry.

\subsection{Two ways to understand soft modes from bulk}\label{eader}

The soft modes in the bulk can be understood in two equivalent ways.
\begin{itemize}
\item In the \textbf{\textit{first}} point of view (left panel in Fig. \ref{fig:my_label}), 
\begin{enumerate}
\item The metric is Poincare $AdS_2$ \eq{ads} (or $AdS_2$ black hole \eq{2dbtz} for finite temperature).
\item The boundary curve is given by $(\tilde \tau,\tilde z)=(f(u), z(u))$. In order to fix the proper length of the boundary we impose $ \gamma_{uu} = l^2/\delta^2$, where $\gamma_{uu}$ is the induced metric on the boundary and $u$ is the boundary time variable. Imposing this gauge we get, $z(u)= \delta\ f'(u) + O(\delta^3)$
\item On the boundary curve $(f(u),\delta\ f'(u))$ we impose $\Phi|_{\del M} = \Phi (f(u), \delta\ f'(u))= \Phi_r(u)/\delta  + O(\delta)$ where $\Phi_r(u)$ is a given boundary condition for the dilaton. Note that the parametric boundary curve is not uniquely described by $(f(u),\delta\ f'(u))$ as we still have a freedom to redefine the boundary time variable $(u(\tilde u))$ however the boundary condition on the dilaton removes this freedom.
\item The path integral is over all possible shapes of the boundary parameterized by $f(u)$, which is the ``Schwarzian'' mode.
\item The Schwarzian mode at finite temperature is parameterized by $f(u) = (\beta/\pi) \tan(\pi \varphi(u)/\beta)$ where $\varphi(u)$ belongs to ${\it Diff}\ \mathbb{S}^1$ group. The length of the domain and the target circle is $\beta$. Note that with this map the $\beta \to \infty$ (zero temperature) limit is well defined. 
\end{enumerate}
\item In the \textbf{\textit{second}} point of view (right panel in Fig. \ref{fig:my_label}), 
\begin{enumerate}
\item The metric is asymptotically AdS$_2$ \eq{aads2} in Fefferman-Graham gauge \cite{Mandal:2017thl}. This metric depends on an arbitrary function $f(\tau)$ and satisfies \eq{rp2}.
\item The boundary curve is given by $(\tau(u),\delta)$ where $u$ is the boundary time variable. Imposing the condition $ \gamma_{uu} = l^2/\delta^2$ tells us $\tau(u)=u$. Hence the boundary curve is $(\tau,z)=(u,\delta)$.
\item On the boundary curve $(\tau,z)=(u,\delta)$, impose $\Phi|_{\del M} = \Phi (u, \delta)= \Phi_r(u)/\delta  + O(\delta)$ where $\Phi_r(u)$ is a given boundary condition for the dilaton.
\item The path integral is over all the metrics \eq{aads2} parameterized by $f(\tau=u)$, which is the ``Schwarzian'' mode. These metrics are obtained by performing large diffeomorphism \eq{eq:exact-large-diff} on \ads geometry.  
\item A geometry with horizon (finite temperature) is parameterized by $f(u) = (\beta/\pi) \tan(\pi \varphi(u)/\beta)$ where $\varphi(u)$ belongs to ${\it Diff}\ \mathbb{S}^1$ group. The length of the domain and the target circle is $\beta$. Note that with this map the $\beta \to \infty$ (zero temperature) limit is well defined. 
\end{enumerate}
\end{itemize}
Metric \eq{aads2} is a large diffeomorphism of the Poincare $AdS_2$ metric \cite{Mandal:2017thl}.

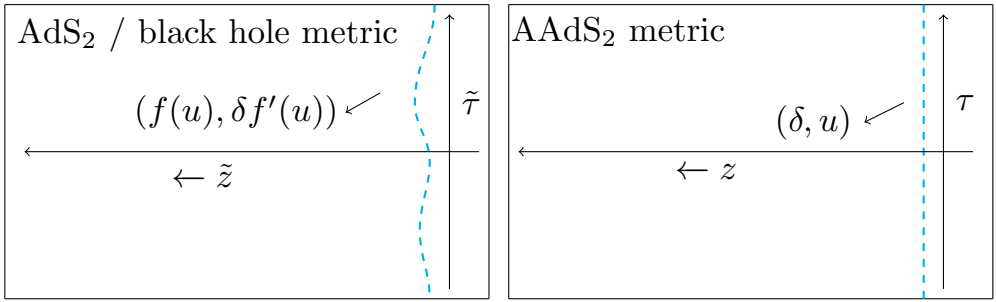
\begin{figure}[h]
    \centering
    \begin{tikzpicture}[scale = 1.3, every node/.style={scale=1.3}]
    \draw (0,0)--(0,3);
    \draw (0,3)--(4.9,3);
    \draw (5.1,3)--(10,3);
    \draw (10,3)--(10,0);
    \draw (10,0)--(5.1,0);
    \draw (4.9,0)--(0,0);
    \draw (4.9,3)--(4.9,0);
    \draw (5.1,3)--(5.1,0);
    \node [below right] at (0,3){AdS$_2$ / black hole metric};
    \node [below right] at (5,3){AAdS$_2$ metric};
    
    \draw [->](4.5,0.1) -- (4.5,2.9);
    \draw [->](4.8,1.5) -- (.2,1.5);
    \draw [cyan, dashed ,  thick] (9.3,3) -- (9.3,0);
    \node [below] at (2,1.5) {$\xleftarrow{} \tilde z$};
    \node [right] at (4.5,2) {$\tilde \tau$};
    
    \draw [->](9.5,0.1) -- (9.5,2.9);
    \draw [->](9.8,1.5) -- (5.2,1.5);
    \draw [thick, cyan,dashed] (4.35,3) to [out=270,in=90] (4.15,2)
     to [out=270,in=90] (4.3,1.4) to [out=270,in=90] (4.2,0.7) to [out=270,in=90] (4.3,0);
    \node [below] at (7.1,1.5) {$\xleftarrow{}  z$};
    \node [right] at (9.5,2) {$\tau$};
    
    \draw [->](3.8,2.1)--(3.45,1.9);
    \node [left]at (3.5,1.9) {$(f(u), \delta f'(u))$};
    
    \draw [->](9.1,2)--(8.7,1.8);
    \node [left] at (8.7,1.8) {$(\delta,u)$};
    \end{tikzpicture}
    \caption{{The picture in the right panel (asymptotically AdS$_2$ metric with straight boundary) is achieved by performing large diffeomorphisms \eq{eq:exact-large-diff} on the left panel (AdS$_2$/\ads black hole metric with wiggly boundary)}}
    \label{fig:my_label}
\end{figure}
Hereafter we will be working with the first point of view, but rewriting following calculations in the second point of view is straight forward.
\subsubsection*{Explicit symmetry breaking terms in $S_{2D}$}

We can write the 2D action\eq{full-action-2d} into 2D bulk and 1D boundary terms as 
\begin{equation}\label{parts}
 S_{2D} = S_0 + S_{\rm{b}},
\end{equation}
This first term $S_0$ is precisely zero in either of two points of view discussed above. The non-zero contribution comes from the boundary part ($S_{\rm b}$) of the  action either in the wiggly boundary picture (first point of view) or using the large diffeomorphisms (second point of view). This is exactly analogues to effective action for pseudo Nambu-Goldstone modes occurring due to explicit symmetry breaking (as described in the beginning of the section). Therefore, we identity $S_b$ with $\Delta S$ of \eq{SpNG}. We continue our further discussion with the first point of view, where the bulk metric is the \ads black hole and the boundary is defined to be $(\tau(u), z(u))=(\L(u), \delta~\L'(u))$. 

The two  terms in \eq{parts} are,
\begin{equation}
 S_0 =  -\frac{1}{16 \pi G_2} \int_{M_2} d\tu dz\ \sqrt{g}\ \Phi \left(R + \frac{2}{l^2} -\frac{l^2}{4} \Phi^2 \tilde{F}^2\right) -\f{i kl}{2} \int_{M_2} d\tu dz\ \s{g}\ \rchi \left( J_0\ \Phi -F\right) \hspace{.85cm}
\end{equation}
and 
\begin{eqnarray}
S_b= \Delta S &=&  \frac{i kl}{4}\int_{\del M_2} d u\ \rchi A_u - \f{k l}{8} \int_{\del M_2} d u ~\Phi\ \sqrt{\gamma}\ \gamma^{u u}\left( A_u^2+ \gamma_{u u}\left(\frac{\rchi}{\Phi}\right)^2 + \left(l\rchi \Phi B_u\right)^2 -2 l\rchi B_{u}A_u\right)\nn \\ 
&&- \frac{1}{8 \pi G_2} \int_{\del M_2} d u \sqrt{\gamma}\ \Phi\ \left(\mathcal K-\f 1 l\right)
\end{eqnarray}
Note that the terms in $S_0$ are invariant under large gauge transformations and large diffeomorphisms which constitute ${\it\ Diff} \times$ U(1)$_{\text{local}}$ but the terms in $S_b$ are not invariant. 

\subsubsection{Boundary behavior of the fields}
{As we discussed, the action for the soft modes is $\Delta S$}. In order to compute it we need the expressions for the fields at the boundary. {Recall that in the reference black hole metric \eq{2dbtz}} the boundary curve is defined by $(\tau(u),z(u)) = (\varphi(u), \delta\ \varphi'(u))$. With this, we outline the boundary behaviour of the fields below. 
\begin{itemize}
\item The \textit{gauge condition} on the \textit{induced boundary metric} is,
\begin{equation}
 \gamma_{uu} = g_{\mu\nu} \left(\frac{d x^\mu}{du}\right) \left(\frac{d x^\nu}{du}\right) = \frac{l^2}{\delta^2} \label{gammauu}
\end{equation}
\item We choose the \textit{boundary condition} on the \textit{dilaton} as 
\begin{equation}\label{phir}
 \Phi|_{\del M_2} = \f{\Phi_r(u)}{\delta} + O(\delta),
\end{equation}
where $\Phi_r(u)$ is arbitrary external field of mass dimension $-1$.
\item The \textit{boundary condition} for \textit{scalar} $\rchi$ and the \textit{KK gauge field} $B_u$ will be $0$.
\end{itemize}
In addition we can show that,
\begin{itemize}
\item the value of the \textit{extrinsic curvature} on the boundary with AdS$_2$ black hole \eq{2dbtz} as the metric is 
\begin{equation}
 \mathcal{K} = 
 \f{1}{l} + \f{\delta^2}{l}\ \bigg\{\tan\left(\f{\pi}{\beta}\ \varphi(u)\right),u\bigg \} + O(\delta^4) \label{K}.
\end{equation}
\item We can also find the behaviour of the induced \textit{boundary gauge field} $A_u$ as follows.\\ The equation \eq{eq:trunc} on the boundary is, 
\begin{equation}\label{bdge}
 F|_{\del M_2} = J_0\ \Phi|_{\del M_2}~.
\end{equation}
In  the gauge $A_z = 0 $, a general form of the field $A_\tau$ can be written down as,
\begin{equation}\label{genat}
 A_{\tau} (\tau,z) = \f{C(\tau)}{z^2} + B + D(\tau)\ z^2 + \dots
\end{equation}
The $z$ independent part of the above expression (i.e., $B$) is made time independent by doing a time dependent residual gauge transformation.\\
 Using \eq{bdge} and \eq{genat} we get
\begin{eqnarray}\label{eq1}
\f{2 C(\varphi(u))}{l^2 \delta \ \varphi'(u)} + O(\delta) &=&  J_0\ \f{\Phi_r(u)}{\delta} + O(\delta)\\
\implies  \f{C(\varphi(u))}{\varphi'(u)} &=& \f{l^2 J_0\ \Phi_r(u)}{2} ~.
\end{eqnarray}
 We have found that the boundary condition on the dilaton \eq{phir} and Eq. \eq{bdge} impose a condition on the leading order term of $A_\tau$ in \eq{genat}. This  condition on the leading order coefficient $C(\tau)$ will be very crucial in the next section where we propose counter terms to keep on shell action finite. 
Next, it is straight forward to define the boundary \textit{gauge field} $A_u$ as,
\begin{equation}\label{Au}
 A_u(u) = A_\tau(\tau(u),z(u)) \left(\frac{d \tau}{du}\right) + A_z(\tau(u),z(u)) \left(\frac{d z}{du}\right).
\end{equation}
Using the relevant expressions, we obtain,
\begin{eqnarray}
 A_u (u) &=& \left(\f{C(\varphi(u))}{(\delta ~ \varphi'(u))^2} + B+ O(\delta^2)\right) \varphi'(u)  \nonumber\\
A_u(u)&=& \f{l^2 J_0\ \Phi_r(u)}{2\ \delta^2} + B\ \varphi'(u) + O(\delta^2).
\label{Aufin}
\end{eqnarray}
\end{itemize}
\subsubsection{Counter terms}\label{sec:ct}
Now that we have all the ingredients to compute the action for the soft modes, it is easy to see that the $S_b$ with fields behaving as above  diverges as $\delta \to 0$ (the on-shell action is not renormalized). In order to see this, let us work with the solutions $\rchi=0$, $B_u =0$ and the AdS$_2$ black hole as the metric. With this choice the bulk terms vanish and we get, 
\begin{equation}\label{eq:total-delta-s}
  S_b = -\f{k l}{8} \int_{\del M_2} d u~\sqrt{\gamma}\ \Phi\  \gamma^{u u} A_u A_u 
- \frac{1}{8 \pi G_2} \int_{\del M_2} d u \sqrt{\gamma}\ \Phi\ \left(\mathcal K-\f 1 l\right)
\end{equation}
We can see that the quadratically divergent term in \eq{Aufin} will make this action divergent \footnote{The reader might wounder, how such a divergence can be understood in 3D which consists of Einstein-Hilbert plus Chern-Simons action but the divergence arises due to coupling to the external source $J_0$}. This divergence can be removed by adding local boundary gauge and diffeomorphism invariant counter terms. We propose the following counter term for the same.
\begin{equation}\label{eq:ct}
S_{CT}=  \f{k J_0 l^2}{32} \int_{\del M_2} du\ \Phi^2 A_u + \f{k}{8J_0} \int_{\del M_2} du\ A_u (\gamma^{uu} A_u A_u)  
\end{equation}
See Appendix \ref{ct} where we give details of the 3D origin of the above terms. The action for pseudo Nambu-Goldstones in presence of these counter terms becomes,
\begin{equation}
  S_{pNG} =   S_b^{Gravity} +  S_b^{Gauge} 
\end{equation}
\begin{align}
 S_b^{Gravity} & =- \frac{1}{8 \pi G_2} \int_{\del M_2} d u \sqrt{\gamma}\ \Phi\ \left(\mathcal K-\f 1 l\right)\nn \\
 S_b^{Gauge} &= -\f{k l}{8} \int_{\del M_2} d u ~\sqrt{\gamma}\ \Phi\  \gamma^{u u} A_u A_u + \f{k J_0 l^2}{32} \int_{\del M_2} du\ \Phi^2 A_u + \f{k}{8J_0} \int_{\del M_2} du\ A_u (\gamma^{uu} A_u A_u)
 \label{deltas}
\end{align}
In the above, we have presented counter terms for the gauge action explicitly (our action in 3D has an additional coupling term  \eq{coupling} with the usual Chern-Simons action). The counter term in the pure gravity action is the well known $-1/l$, as used in \eq{S3Dgrav}. This does not need any special treatment in this work.
\subsection{The soft mode effective action}
\label{sec:schwarzian}
Drawing parallel with Section \ref{eader}, the effective action from $\Delta S_{Gravity}$ can be evaluated using \eq{gammauu}, \eq{phir} and \eq{K}. 
\begin{align}\boxed{
S_b^{Gravity}  = - \f{1}{8 \pi  G_2} \int du\  \Phi_r (u)\ \bigg\{\tan\left(\f{\pi\ \varphi(u)}{\beta}\right),u \bigg \} }
\label{schwarzian}
\end{align}
Note that in the gravity path integral is an integral over different boundary shapes given by $\varphi(u)$. We have the AdS$_2$ black hole geometry \eq{2dbtz} and the boundary condition \eq{phir} hence in the above effective action $\varphi(u)$ is the dynamical variable and $\Phi_r$ is an external coupling.\vskip.1cm
The $U(1)$ soft mode is the residual gauge transformation which is only a time dependent function as we are working in the $A_z =0$ gauge. Therefore, this gauge transformation acts as follows
\begin{equation}
 A_\tau(\tau,z) \to A_{\tau}(\tau,z) + \del_\tau \tilde \phi(\tau)
\end{equation}
with this and \eq{Au}, $A_u$ will transform as follows
\begin{equation}\label{u1sm}
 A_u(u) \to A_u(u) + \del_{\tau}\tilde \phi(\varphi(u)) \varphi'(u) = A_u(u) + \del_u\phi(u) 
\end{equation}
where $\phi(u) = \tilde \phi(\varphi(u))$ is the $U(1)$ soft mode. Thus the soft mode action $\Delta S_{Gauge}$, evaluated using \eq{gammauu}, \eq{phir}, \eq{Aufin} and \eq{u1sm} is
\begin{equation}\boxed{
 S_b^{Gauge} = \frac{k}{16}\ \int d u\ \Phi_r(u)\ (\del_u \phi(u) + B\ \varphi'(u) )^2.
 \label{sigma-model}}
\end{equation}
where $B$ is an undetermined constant in \eq{Aufin} which can not be fixed by \eq{bdge}. 
\subsection{Combined effective action}
The combined effective action given by adding \eq{schwarzian} and \eq{sigma-model} is 
\begin{equation}
 S_{pNG} =  \int du\  \Phi_r(u) \left(\frac{k}{16}\ (\del_u \phi(u) + B\ \varphi'(u) )^2 - \f{1}{8 \pi  G_2}\ \bigg\{\tan\left(\f{\pi\ \varphi(u)}{\beta}\right),u \bigg \}\right)~,
\end{equation}
here $\L(u)$ and $\phi(u)$ are dynamical variables whereas $\Phi_r(u)$ is an external coupling which can be made constant by choosing a new boundary time $\Phi_r(u) (d\tilde u /du) =  \bar \Phi_r $\cite{Maldacena:2016upp}. 
\begin{equation}\boxed{
 S_{pNG} =  \int_0^\beta d \tilde u\ \left(\frac{\bar \Phi_r k}{16}\ (\phi'(\tilde u) + B\ \L'(\tilde u) )^2 - \f{\bar \Phi_r\ }{8 \pi  G_2}\ \bigg\{\tan\left(\f{\pi}{\beta}\ \L(\tilde u)\right),\tilde u \bigg \}\right)~
 \label{beffac}}
\end{equation}
This is the effective action for the reparameterization and the $U(1)$ soft mode. To determine B, we compare the  on-shell action for the black hole solutions \eq{2dbtz}, \eq{btzgendil1} and \eq{atgen} with the on-shell action computed from \eq{beffac} with the corresponding soft modes. Plugging these solutions in $\Delta S_{Gauge}$ \eq{deltas} with the identification, $\bar \Phi_r = a$,  we get,
\begin{equation}\label{os1}
 S_{b_{on-shell}}^{{Gauge}} =  \f{\bar \Phi^3_r\ k\ l^4\ \pi^4\ J_0^2 }{16 \beta^4}\ \beta.
\end{equation}
The corresponding solution as soft modes are $\L(\tilde u) = \tilde u$ and $\phi(\tilde u)=$ constant. Plugging these in \eq{sigma-model} we get
\begin{equation}\label{os2}
S_{b_{on-shell}}^{Gauge} = \f{\bar \Phi_rk}{16}\ B^2\ \beta.
\end{equation}
Comparing \eq{os1} and \eq{os2} we get
\begin{equation}
 \boxed{B = -\ \f{\Phi_rJ_0l^2\pi^2}{\beta^2}}
\end{equation}
(The sign is fixed by comparing to black hole solution directly).
When we compare the action \eq{beffac} with the field theory effective action in the next subsection we will see that $B$ will correspond to the chemical potential of the field theory and as pointed out earlier, it is proportional to the coupling $J_0$.  
\subsubsection*{$ S_b^{Gauge}$ effective action with $B =0$}
We can consider the case where $B$ is 0, this is equivalent to putting $J_0=0$. From \eq{eq:trunc}, it implies that there no electric flux in the system. This case is equivalent to putting the renormalized boundary value of the gauge field as pure gauge. The action is dual to $U(1)$ charged SYK model without chemical potential and the effective action is 
\begin{equation}
 S_b^{Gauge} = \frac{k}{16}\ \int du\ \Phi_r(u)\ \left(\del_u \L(u)\right)^2.
\end{equation}
these results have recently appeared in \cite{Mertens:2018fds} and \cite{Gonzalez:2018enk}.
\subsection{Comparison with field theory}
Rewriting  $\varphi(\tilde u) = \tilde u + \ep(\tilde u)$ and ignoring surface terms plus constants from the action we get
\begin{equation}\label{our-result}
 S = \int_0^\beta d \tilde u\ \bar \Phi_r\ \left(\frac{k}{16}\ (\phi'(\tilde u) + B\ \ep'(\tilde u) )^2 - \f{1}{8 \pi  G_2}\ \bigg\{\tan\left(\f{\pi}{\beta} (\tilde u + \ep(\tilde u))\right),\tilde u \bigg \}\right)~.
\end{equation}
An effective action for the pseudo NG modes of the U(1) charged SYK model was presented in \cite{Davison:2016ngz} (where we have substituted chemical potential using  definitions in Appendix B of \cite{Davison:2016ngz}),
\begin{equation}\label{subir}
 S = N \int_0^\beta d \tilde \tu\ \left( \frac{K }{2}\ (\del_{\tilde \tau} \phi(\tilde \tau) - i\ \tilde \mu \ \ep'(\tilde \tu) )^2 - \f{\gamma}{4 \pi^2}\ \qty{{\rm tan} \left({\pi\ (\tilde \tu + \ep(\tilde \tu))}/{\beta}\right),\tilde \tau} \right)
\end{equation}
here $K$ and $\gamma$ are  field theory thermodynamic parameters determining compressibility and specific heat respectively. In the large $\mathcal{J}$ (where $2 \mathcal J^2 = q^2 J^2/(2+2 \cosh(\mu/T))^{q/2-1}$), small $\mu$ (near conformal limit) and large q limit, $K = q^2 / (16 \mathcal{J})$ and $\gamma = 2\pi^2 / (\mathcal{J}\ q^2)$. Our result \eq{our-result} matches with \eq{subir} with following identifications
\begin{equation}
 \bar \Phi_r = \f{1}{\mathcal{J}}, \qquad B = -\ \f{\Phi_rJ_0l^2\pi^2}{\beta^2} = - i \tilde \mu, \qquad \f{N q^2}{2} = k, \qquad \f{ 4\pi N}{q^2} = \f{1}{ G_2}.
 \label{parameters}
\end{equation}
\section{Quantum chaos \label{sec:chaos}}
So far our discussion has mostly focused on the soft modes of the charged SYK model and their bulk dual. As in case of the uncharged model, there are other modes represented by operators of the form $\psi_i\!\stackrel{\leftrightarrow}{\del_\tau^p}\!\psi_i$,
$\psi^\dagger_i\!\stackrel{\leftrightarrow}{\del_\tau^p}\!
\psi_i^\dagger$, and
$\psi^\dagger_i\!\stackrel{\leftrightarrow}{\del_\tau^p}\!\psi_i$.
Here the two-sided arrow schematically represents a specific combination of terms with $n$ derivatives on the first fermion and $p-n$ derivatives on the second fermion. The first two types of operators are charged while the third set are neutral.  The story of correlators of the neutral operators is similar to that in the uncharged SYK model (briefly reviewed below in Sec \ref{sec:neutral-OTO}). Therefore, in what follows, we will be concerned with charged operators.  Let us denote a typical pair of conjugate charged operators by $O(\tau)$, $O^*(\tau)$, with scaling dimension $\Delta$.\footnote{In this section, we will reserve the   notation $\Delta$ for the scaling dimension of $O$, rather than for  that of the fermion.}

As in \cite{Mandal:2017thl, Maldacena:2016upp}, to compute correlators of these operators from the bulk viewpoint we consider probe scalar fields $\eta(z,\tau)$, $\eta^*(z,\tau)$ whose masses are given by the standard AdS/CFT correspondence between mass and operator dimensions. The classical action for the probe scalars will be given by 
\begin{align}
S_{matter} =  \f1{16\pi G_N}  \int \sqrt{g}
\left[
\left(g^{\al\be} D_\al \eta^* D_\be \eta +  m^2 |\eta|^2\right)
+ ... \right]
\label{charged-matter}
\end{align}
where the ... terms are higher order in the fields $\eta$. Here the
mass $m$ of the scalar field is determined by the dimension $\Delta$
of the dual operator $O$:
\begin{align}
  \Delta = 1/2 + \nu, \; \nu= \sqrt{1/4 + m^2 l^2}
  \label{mass-dim-2D}
  \end{align}
In Appendix \ref{app:chaos-3D} we have considered a probe scalar action
\eq{charged-matter-dilaton} which is coupled to the Dilaton and is more
natural from the 3D viewpoint; we show there that the main conclusions
of the remainder of this section remain unchanged--- the main change
being in the mass dimension formula above, which changes to
a different one \eq{mass-dim-hybrid}.

\subsection{OTO correlators in the uncharged model\label{sec:neutral-OTO}}
To begin, we will quickly review the calculation in the uncharged case. Let us consider the Euclidean correlator $\lan O(\tau_1)
O(\tau_2) O(\tau_3) O(\tau_4) \ran$, which is given by
\begin{equation}
    \langle \mO(\tau_1) \mO(\tau_2)\mO(\tau_3)\mO(\tau_4) \rangle =
    \left. \frac1{Z[J]} \frac\de{\de J(\tau_1)}\frac\de{\de
      J(\tau_2)}\frac\de{\de J(\tau_3)}\frac\de{\de J (\tau_4)} Z[J]
    \right|_{J=0}
\label{peskin}
\end{equation}
By the usual rules of AdS/CFT, the quantity $Z[J]$ is given by a bulk path integral with the following non-normalizable (source-type) boundary condition over the dual scalar field $\eta$ 
\begin{equation}
    \eta(\tau,z) = z^{\frac12-\nu} J(\tau) +
 \left(\hbox{higher order in}~z \right),
\label{eta-bc}
\end{equation}
in addition to the non-normalizable dilaton boundary condition  we described in the previous sections. We will (a) first compute the matter path integral, for a fixed metric characterized by a particular Diff element $\L$, and then (b) integrate over the metrics (i.e. integrate over the functions $\L$;  note that as pointed out earlier this is equivalent to fixing the metric and integrating over different boundary cut outs). This procedure gives us
\begin{align}
& Z[\L,J] =  \exp[\int d\tau d\tau' J(\tau)
G^{\L}_{\beta,\De}(\tau,\tau')J(\tau')],
\quad 
G^{\L}_{\beta,\De}(\tau,\tau') \equiv 
\bqty{ \L'(\tau) \L'(\tau') }^\De G_{\beta,\De}(\L(\tau)-\L(\tau'))  
\label{z-f}\\
& Z[J]=  \int \qty[d\mu[\L] ] \exp[S_{\rm eff}[\L[\tau]]] Z[\L,J]
\label{f-path-integral}
\end{align}
Here $d\mu[\L] \equiv \prod_\tau\f{\mathcal{D} \L(\tau)}{\L'(\tau)}$ is an \slr-invariant measure \cite{Mandal:2017thl,Stanford:2017thb}.  The effective action $S_{\rm eff}[\L]$ is given by \eq{schwarzian}. To derive the first line, note that if we consider $\L(\tau)=\tau$ (the identity transformation), corresponding to the AdS$_2$ black hole metric \eq{2dbtz} (or the $(\tau,z) = (\tau,\delta)$ boundary curve according to the first point of view in section \ref{sec:soft-mode}), we will have the familiar result (using tilde coordinates to distinguish the starting point of the orbit of large diffeomorphism, which is AdS$_2$ black hole.)
\[
Z[\L=\tau,J] =  \exp[\int d\tilde\tau d\tilde\tau' 
\tilde J(\tilde\tau) G_{\beta,\De}(\tilde\tau-\tilde\tau')  
J(\tilde\tau')]
\]
where $G_{\beta,\De}(\tilde\tau-\tilde\tau')$ is the thermal two-point
function 
\begin{align}
G_{\beta,\De}(\tilde\tau-\tilde\tau')
=\lan O(\tilde\tau) O(\tilde\tau')\ran
=\left[\f{\pi}{{\beta} \sin(\f{\pi(\tilde\tau
- \tilde\tau')}{\beta})}\right]^{2 \De}
\label{thermal-2-pt}
\end{align}
To compute $Z[\L,J]$ we need to do this computation in the metric \eq{aads2} (with $f(\tau) = (\beta/\pi) \tan(\pi \L(\tau) /\beta)$) and \eq{eq:generalPhi}. This can be done by applying a large diffeomorphism transformation asymptotically and
noting that under such a transformation the source term $J$ transforms
as 
\[
{\tilde z}^{\f12-\nu} \tilde J(\tilde \tau) =z^{\f12-\nu} 
J( \tau)
\]
by virtue of \eq{eta-bc} and the fact that $\eta$ is a scalar. This leads to the expression for $G^{\L}_{\beta,\De}$ in \eq{z-f}.

Eq. \eq{f-path-integral} is simply obtained by noting that the integral over the space of metrics, as shown in \cite{Mandal:2017thl}, reduces to the integration of the $\L$-variables with the Schwarzian effective action. 

Using \eq{peskin}, \eq{z-f}, \eq{f-path-integral} we get

\begin{align}
   & \langle \mO(\tau_1) \mO(\tau_2)\mO(\tau_3)\mO(\tau_4) \rangle 
\nonumber\\
& = \int \qty[ d\mu[\L]] \ \exp[S_{\rm eff}[\L]] \Bigg(
G^{\L}_{\beta,\De}(\tau_1, \tau_2)\ G^{\L}_{\beta,\De}(\tau_3, \tau_4)
    + (\tau_2\leftrightarrow \tau_3) + (\tau_2\leftrightarrow \tau_4) \Bigg) 
\label{wick}
\end{align}

In the $G_N \to 0$ limit, it is enough to expand $S_{\rm eff}[\L]$ up to quadratic order in $\ep$ defined by $L(\tau) = \tau + \sqrt{16\pi
  G_N}\ \ep(\tau)$:
\[
S_{\rm eff}[\ep]= \f{\bar\Phi_r}{16 \pi G_2}
\int_0^\be d\tau \left[ \left(\ep''\right)^2 - 
\left(\f{2\pi}{\be}\right)^2  \left({\ep'}\right)^2 \right] + ...
\]
with propagator 
\begin{align}
D(\om)=\left(\f{8\pi G_2}{\bar\Phi_r}\right)\ \f1{\om^2\left[\om^2- (2\pi/\be)^2\right]}
\label{ep-prop}
\end{align}
Correspondingly,in \eq{wick}, we need to expand the Green's
functions to leading order in $\ep$; thus: 
\begin{align}
&G^{\L}_{\beta,\De}(\tau_1, \tau_2)= G_{\beta,\De}(\tau_1, \tau_2) +
\ep(\tau_1)\delta G_1 + \ep(\tau_2)\delta G_2 + O(\ep^2),
\nonumber\\
&G^{\L}_{\beta,\De}(\tau_3, \tau_4)= G_{\beta,\De}(\tau_3, \tau_4) +
\ep(\tau_4)\delta G_3 + \ep(\tau_4)\delta G_4 + O(\ep^2).
\label{g-f-expansion}
\end{align}
Collecting all, the Euclidean four-point correlator is of the form
\begin{align}
   & \langle \mO(\tau_1) \mO(\tau_2)\mO(\tau_3)\mO(\tau_4) \rangle 
= G_{\beta,\De}(\tau_1, \tau_2)\ G_{\beta,\De}(\tau_3, \tau_4)
    + (\tau_2\leftrightarrow \tau_3) + (\tau_2\leftrightarrow \tau_4)
\nonumber\\ 
& + \f{1}{16\pi G_N} \Bigg(\delta G_2\ \delta G_3 \lan 
\ep(\tau_2)\ep(\tau_3) \ran + 
\left[(2,3) \rightarrow (1,3)
\right]+ \left[(2,3)  \rightarrow(1,4)
\right]+ \left[(2,3) \rightarrow  (2,4)
\right] \Bigg).
\label{wick-ep}
\end{align}
In the same manner as in the boundary theory calculation\cite{Maldacena:2016hyu}, an OTO correlator $\lan O(0) O(T) O(0) O(T) \ran$ can be obtained by an appropriate analytic continuation of the above expression. The chaotic growth originates from the $\ep$-propagators in the second line of the above equation (see the schematic representation in Figure \ref{fig-boundary-graviton} of the first term in parenthesis in the above equation). The pole $\om = -2\pi/\be$ of the propagator\eq{ep-prop}, under the Lorentzian continuation leads to a growing term proportional to $\exp[\l_L T]$ for large $T$, with $\l_L =2\pi/\be$.\footnote{The way this happens in practice is a bit more  subtle\cite{Maldacena:2016hyu}; the poles of \eq{ep-prop}, $\om = 0,  \pm 2\pi/\be$, correspond to \slr zero modes, and hence are excluded  from the Matsubara sum involved in the real time propagator $\lan \ep(0) \ep(\tau)\ran$. However, the sum over all other frequencies   leads to a contour integral which gets deformed to a new contour  which includes the only the poles $\om = 0, \pm 2\pi/\be$.
\label{ftnt:sl2r-zero-modes}}
\begin{center}
\begin{tabular}{cc}
\begin{minipage}{0.45\hsize}
\begin{figure}[H]
\centerline{\includegraphics[height=5cm]{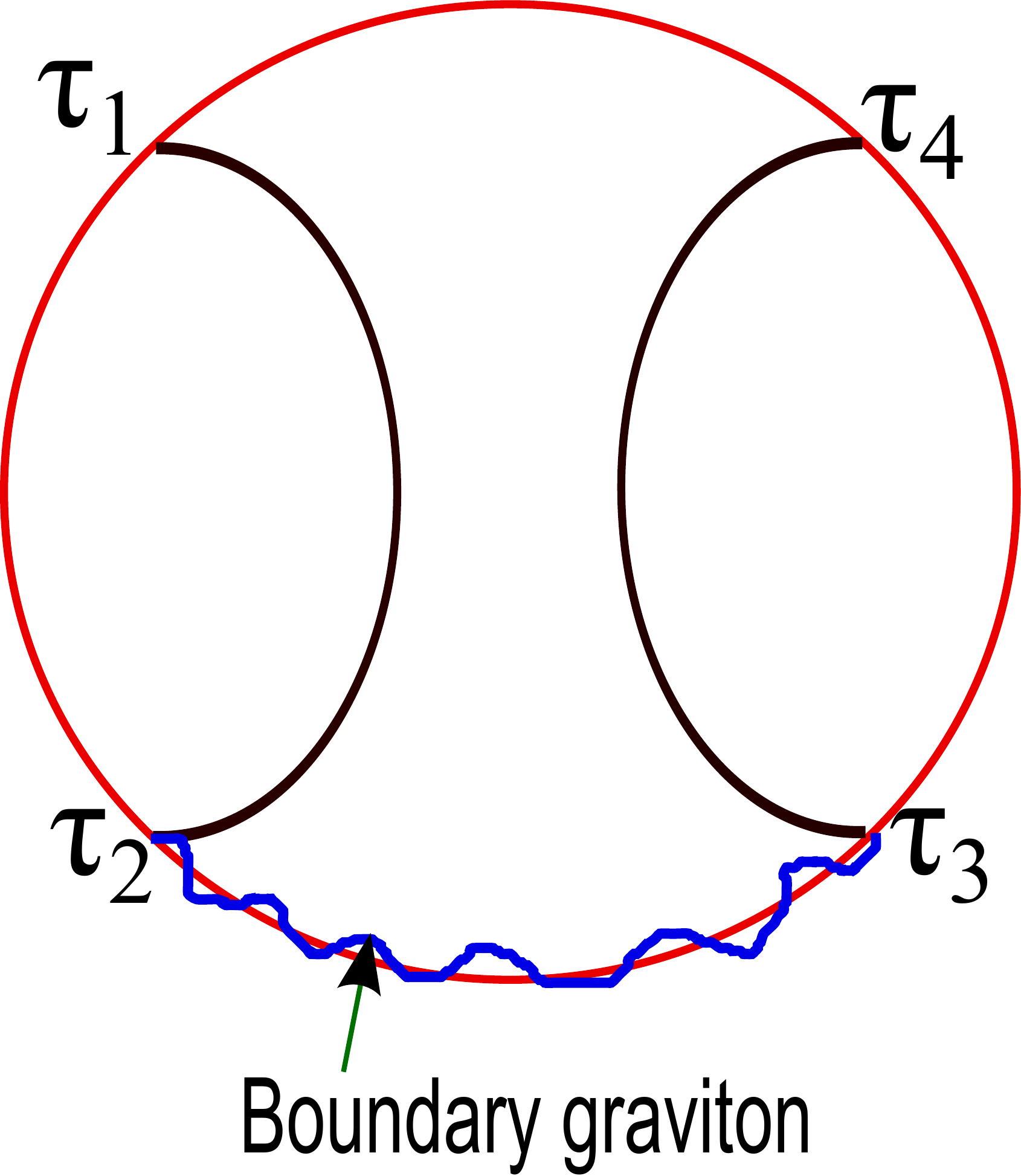}}
\caption{\footnotesize {Bulk description of OTO correlators in the
    uncharged SYK model involves exchange of boundary gravitons; the
    graviton propagator \eq{ep-prop} has a pole $\om= -2\pi/\b$ which
    corresponds to maximal Liapunov exponent $\l_L = 2\pi/\b$.  } }
\label{fig-boundary-graviton}
\end{figure}
\end{minipage}
\kern15pt
\begin{minipage}{0.45\hsize}
\begin{figure}[H]
\centerline{\includegraphics[height=5cm]{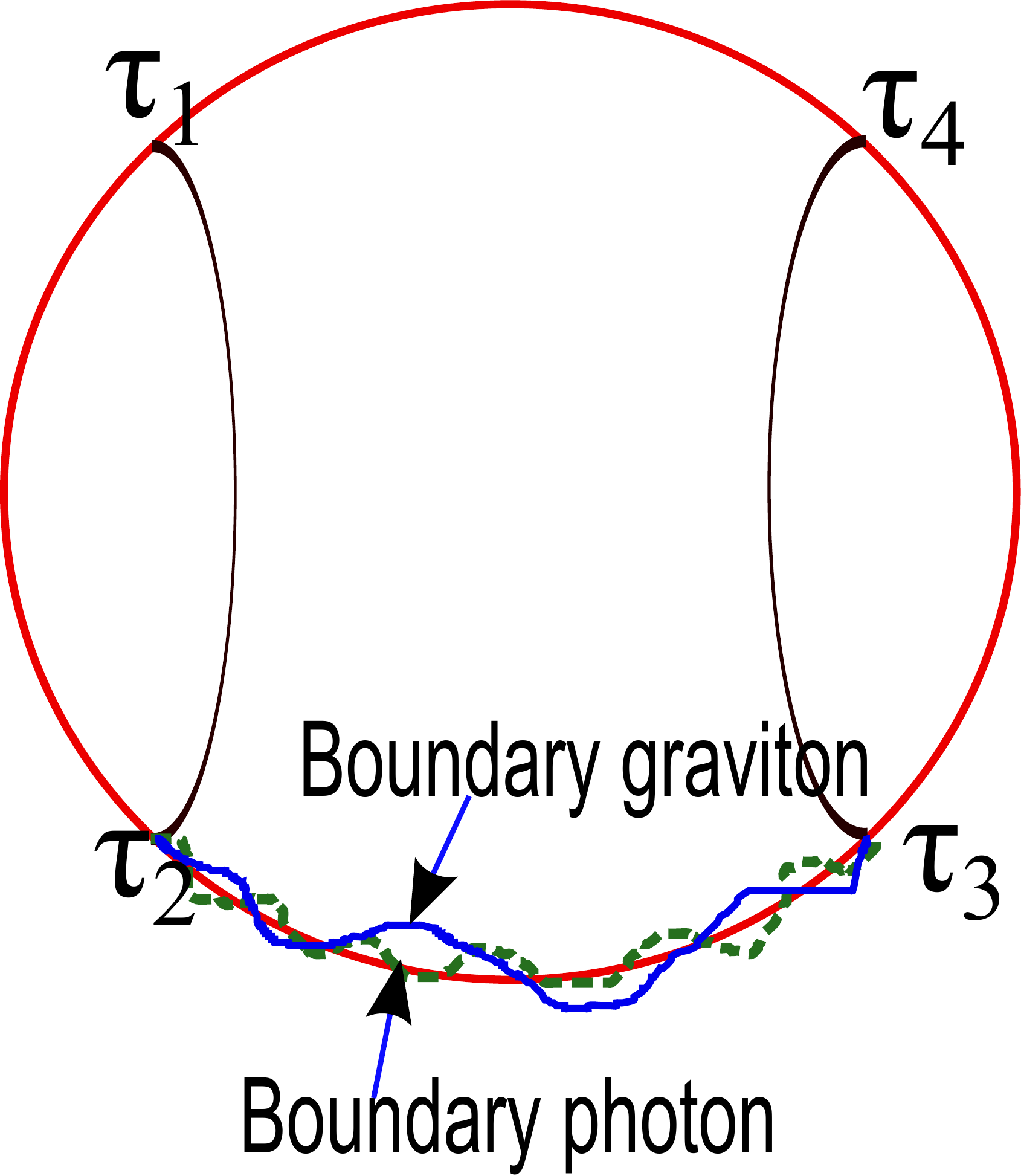}}
\caption{\footnotesize {For the charged SYK model, the bulk
    description of OTO correlators of charged operators involves
    exchange of boundary gravitons and as well as of boundary
    photons. After appropriate rediagonalization, the gravitons and
    photons are decoupled. The photons do not have any nontrivial pole
    and have zero Liapunov exponent.} }
\label{fig-boundary-photon-graviton}
\end{figure}
\end{minipage}
\end{tabular}
\end{center}
\subsection{OTO correlators in the charged model}
\label{sec:charged-OTO}
Let us now come back to the charged model.  We would like to compute, from the bulk dual, the OTO correlator of a charged operator $O$ (with charge $q$): $\lan O(0) O^*(t) O(0) O^*(t) \ran$. The computation follows along similar lines as in the uncharged case; we will thus highlight the essential differences.

The bulk path integral is now given by ({\it cf.} 
\eq{z-f}, \eq{f-path-integral}
\begin{align}
& Z[\phi,\varphi,J] =  \exp[\int d\tau d\tau' J(\tau)
G^{\phi,\L}_{\beta,\mu,\De}(\tau,\tau')J(\tau')],
\nonumber\\
&G^{\phi,\varphi}_{\beta,\mu,\De}(\tau,\tau') \equiv 
\bqty{ \L'(\tau) \L'(\tau') }^\De G_{\beta,\mu,\De}(\L(\tau)-\L(\tau')) 
\exp[i q\left(\L(\tau_2) - i \L(\tau_2)\right)] 
\label{z-f-phi}\\
& Z[J]=  \int \qty[d\mu[\phi, \varphi]] 
\exp[S_{\rm eff}[\phi[\tau], \varphi[\tau]]] Z[\phi,\varphi, J]
\label{f-phi-path-integral}
\end{align}
The effective action $S_{\rm eff}$ is now given by \eq{our-result}. The additional part in the expression for $G^{\L,\varphi}_{\beta,\mu,\De}$, involving the large gauge transformation $\phi$, comes from the fact from the gauge transformation of $J$ which is inherited from that $\eta$ ({\it cf.} \eq{eta-bc}). The counterpart of \eq{g-f-expansion} is now given by (up to leading term)
\begin{align}
&G^{\phi,\L}_{\beta,\mu,\De}(\tau_1, \tau_2)= G_{\beta,\mu,\De}(\tau_1, \tau_2) +
\ep(\tau_1)\delta G_1 + \ep(\tau_2)\delta G_2 + 
\phi(\tau_1)\delta G'_1 + \phi(\tau_2)\delta G'_2,
\nonumber\\
&G^{\phi,\L}_{\beta,\mu,\De}(\tau_3, \tau_4)= G_{\beta,\mu,\De}(\tau_3, \tau_4) +
\ep(\tau_4)\delta G_3 + \ep(\tau_4)\delta G_4 +
+ \phi(\tau_3)\delta G'_3 + \phi(\tau_4)\delta G'_4.
\label{g-f-phi-expansion}
\end{align}
Now, \eq{our-result}, {\it a priori}, leads to mixed propagators $\lan \ep \phi \ran$. However, by making a field redefinition $\phi \to \tilde\phi =\phi - i \tilde \mu \ep$, the two fields get decoupled.  This leads to an expression similar to \eq{wick-ep} which involves terms with $\ep$-propagators as well as $\tilde\phi$ propagators. The latter propagator is $\propto 1/\om^2$, and does not have any other nontrivial pole in the complex plane. The structure of the Euclidean propagator is schematically represented in Figure \ref{fig-boundary-photon-graviton}. The Liapunov growth is governed by the graviton propagator, as before; the boundary photon propagator does not show any chaotic growth--- its Liapunov exponent is zero. This result was earlier obtained from a field theory calculation in \cite{Bulycheva:2017uqj}.

\subsection*{Acknowledgements}
We would like to thank Avinash Dhar, Rajesh Gopakumar, Indranil Halder, Arnab Kundu, R. Loganayagam, Shiraz Minwalla, Prithvi Narayan, Pranjal Nayak, Rohan Poojary, Naveen S Prabhakar, Subir Sachdev, Ashoke Sen, Nemani Suryanarayana, Sandip Trivedi and Jungi Yoon for discussions. The results of this paper were first presented by G.M.  \cite{GM:Talk-kyoto-2017} in the Workshop ``Holography and Quantum Dynamics'', held on November 11, 2017 in YITP, Kyoto, Japan; he would like to thank the organizers and participants of this workshop for several fruitful discussions. S.R.W. would like to acknowledge the support of the Infosys Foundation Homi Bhabha Chair at ICTS.

\begin{appendices}

 \section{KK reduction of 3D action}
 \label{appen:KK}
We start with the general KK ansatz,
\begin{equation}
ds^2=g^{(3)}dx^M dx^N= e^{2 \alpha \phi} g^{(2)}_{\mu \nu } dx^\mu dx^\nu+ l^2 e^{2 \beta \phi} (d\theta + B_\mu dx^\mu)^2
\end{equation}
here $\alpha$ and $\beta$ are some constants.  The labels $M$, $N$ run over 3D coordinates  $\{\tau, z, \theta\}$, and $\mu, \nu $ run over 2D coordinates $\{\tau, z\}$. The metric components $g_{\mu \nu}$, scalar and vector fields, $\phi$ and $B_{\mu}$ are independent of the compactified direction $\theta$. Under this identification we obtain, 
\begin{eqnarray}
R^{(3)} &=&e^{-2\alpha \phi}\left[
 R^{(2)}-2 \beta^2 \partial_\mu \phi\partial^\mu \phi-2(\alpha+\beta)~ \nabla^2 \phi - \frac{l^2 e^{-2(2\alpha-\beta)\phi}}{4} \tilde{F}^2\right]~;\nonumber\\
K^{(3)}&=&e^{\alpha \phi}\left(K^{(2)}+\frac{\alpha+\beta}{\beta}e^{-\beta \phi}n^\alpha \nabla_\alpha e^{\beta \phi}\right) \nonumber\\
\sqrt{g^{(3)}} &=& l\ e^{(2\alpha+\beta)\phi} \sqrt{g^{(2)}}~~;~~\sqrt{h} = l\  e^{(\alpha+\beta)\phi} \sqrt{\gamma}
\end{eqnarray}
$\tilde F$ is the field strength tensor for the KK vector field $B=B_{\mu}dx^{\mu}$. Thus the 3D action \eq{S3Dgrav} reduces to a 2D action, 
\begin{align}
 S_{\rm{2DGrav}} = &-\frac{l}{8 G_3} \int_{M_2} d\tu dz\ \left[ \sqrt{g^{(2)}}\ \Phi^{-\beta}  \left(  R^{(2)} + \frac{2}{l^2} -\frac{2}{\Phi^2}(\alpha+\beta+\beta^2) \partial_\mu \Phi\partial^\mu \Phi+2(\alpha+\beta)~\frac{1}{\Phi} \nabla^2 \Phi \right.\right.\nonumber\\
&\hspace{2.5cm} \left.\left.- \frac{l^2 \Phi^{2(2\alpha-\beta)}}{4} \tilde{F}^2\right)\right] - \frac{l}{4 G_3} \int_{\del M_2} d\tu\ \sqrt{\gamma}\ \Phi^{(-2\alpha-\beta)}\left( K^{(2)} - \frac{\alpha+\beta}{\Phi}  n^{\alpha} \nabla_\alpha \Phi-\f 1 l\right) 
\end{align}
where we have redefined $\phi$ as $\phi = -\log\Phi$. Choosing specific values for the constants as $\alpha=0$ and $\beta=-1$,  we can remove the kinetic terms for the dilaton using divergence theorem. This choice of constants leads to, 
\begin{equation}
 S_{\rm{2DGrav}} = -\frac{l}{8 G_3} \int_{M_2} d\tu dz\ \sqrt{g^{(2)}}\ \Phi \left(R^{(2)} + \frac{2}{l^2} - \frac{l^2}{4} \Phi^2 \tilde F^2\right)- \frac{l}{4 G_3} \int_{\del M_2} d\tu \sqrt{\gamma}\ \Phi\ \left(K^{(2)}-1\right)  
\end{equation}
With the identification, $G_3= 2\pi l G_2$, we obtain the action \eq{eq:s2dgrav}.

Under KK reduction we write the 3D CS gauge field $A_M dx^{M} = A_\mu dx^{\mu} + \rchi\  l d\theta$, where $\rchi$ is 2D scalar field. Using metric and gauge field reduction and dropping all the $y$ derivatives, it is straight forward to obtain action \eq{eq:s2dgauge}
\section{Asymptotically \ads uplifted to 3D}\label{app:oxidation}
As seen in section \ref{sec:all-equations}, with the solution $B_\mu=0$ and $\rchi=0$, the general solutions to the equation \eq{phivar} are the  asymptotically \ads spacetimes,  
\begin{equation}
 ds^2_{\rm aads} = \frac{l^2}{z^2} \left( dz^2 + d\tau^2 \left( 1 - \frac{z^2}{2} \qty{f,\tau}\right)^2\right).
\end{equation}
In this background the solution for the dilaton is, 
\begin{equation}
 \Phi= \frac{a+ b\tilde \tau+c(\tilde\tau^2+\tilde z^2)}{\tilde z}~,
\end{equation}
where the $\{\tilde \tau, \tilde z\}$ are given by, 
\begin{align}\
\tilde\tau = f(\tau) - \frac{2 z^2 f''(\tau) f'(\tau)^2}{4 f'(\tau)^2 + 
z^2 f''(\tau)^2}, \quad \tilde z = \frac{4 z f'(\tau)^3}{4 f'(\tau)^2 + 
z^2 f''(\tau)^2}~,
\end{align}
 The 3D oxidation of this solution will be, 
\begin{equation}
\label{eq:uplift}
 ds^2_{\rm uplift} = \frac{l^2}{z^2} dz^2 +\frac{l^2}{z^2} \left( 1 - \frac{z^2}{2} \qty{f,\tau}\right)^2 d\tau^2+l^2\Phi^2 d\theta^2~.
\end{equation}
It satisfies the 3D Einstein equations with $\Lambda=-1/l^2$, see equations \eq{einstein-eq-3D}. This 3D solution violates the Brown and Henneaux conditions. In the metric \eq{eq:uplift} the $g_{\theta \theta}$ component corresponds to a non-normalizable
deformation of AdS$_3$ (parameterized by the $f(\tau)$).

\section{General expression for gauge field in black hole background}
\label{gena}
The \ads black hole metric is
\begin{equation}
 ds^2 = \frac{l^2}{z^2} \left( dz^2 + d\tau^2 \left( 1 - z^2\ \f{\pi^2}{\beta^2}\right)^2\right)~.
\end{equation}
The general solution for dilaton in this background is
\begin{equation}
 \Phi(\tau, z) = (a \pi^2 + c \beta^2 )~ \frac{\left(\beta ^2+\pi ^2 z^2\right)}{2 \pi^2  \beta^2  z } + \left((a \pi^2 - c \beta^2 ) \cos \left(\frac{2 \pi  \tau }{\beta }\right) + b \pi \beta  \sin \left(\frac{2 \pi  \tau }{\beta }\right)\right)~\frac{\beta^2 -\pi^2 z^2 }{2 \pi^2  \beta^2  z }~.
\end{equation}
Using the equation of motion \eq{eq:sourceeq} in the gauge $A_z(\tau,z)=0$ we get,
\begin{equation}
 \del_z A_\tau(\tau,z) +   J_0 \sqrt{g}\ \Phi(\tau, z)= 0~.
\end{equation}
General solution for the above equation is
\begin{equation}
\label{eq:Atau}
 A_\tau(\tau,z) = a_1(\tau)\ \f{1}{z^2} + a_2(\tau)\ \log(\frac{\pi z}{\beta}) + a_3(\tau) z^2 + a_4
\end{equation}
where
\begin{eqnarray}
a_1(\tau) &=& \frac{J_0 l^2}{4 \pi^2 } \left( a \pi^2+ c \beta^2+ (a\pi^2 - c\beta^2)  \cos \Theta + b\ \pi\beta  \sin\Theta \right) \\
a_2(\tau) &=& \frac{J_0 l^2}{\beta^2 } \left((a \pi^2 - c \beta^2) \cos \Theta+b\ \pi\beta  \sin\Theta\right)\hspace{1cm}\\
a_3(\tau) &=& \frac{\pi^2 J_0 l^2}{4 \beta^4}\left(a\pi^2+c\beta^2-(a\pi^2-c\beta^2)  \cos \Theta -b \pi \beta  \sin \Theta\right)\\
a_4 &=& - \frac{J_0 l^2}{2 \beta^2} (a \pi^2 + c \beta^2)
\end{eqnarray}
where $\Theta=2 \pi  \tau /\beta $. We have fixed the integration constant by ensuring that $A_\tau(\tau,z)$ vanishes on the horizon $z = \beta/ \pi$. 
\subsection*{Solutions with $b=0$ and $c = a \pi^2/\beta^2$}
The solution for the dilaton and the gauge field becomes 
\begin{equation}
 \Phi(\tau, z)= \frac{a}{z} + \frac{a \pi^2}{\beta^2}\ z 
\end{equation}
and
\begin{equation}
 A_\tau(\tau,z) = a_1\ \f{1}{z^2} + a_3 z^2 + a_4,
\end{equation}
where
\begin{equation}
a_1(\tau) = \frac{a  J_0 l^2}{2 },\qquad
a_3(\tau) = \frac{a J_0 l^2\pi^4}{2 \beta^4}, \qquad
a_4 = - \frac{aJ_0 l^2 \pi^2}{ \beta^2}.
\end{equation}
These are Eqns, \eq{btzgendil} and \eq{atgen}, which we write here for completeness. Also, note that with this choice of parameters, the solutions become independent of time.

\section{Counter terms}\label{ct}
In this appendix we will discuss the 3D origin of the counter terms proposed in \eq{eq:ct}. We first show that these counter terms cancel divergences in the 3D BTZ background. We then go on to show that they continue to cancel divergences in the 2D theory after the KK reduction. 

\subsection{3D origin of counter terms}
The 3D action for the CS gauge fields with the gravity coupling term is 
\begin{equation}
  S_{\text{3D-Gauge}} = \frac{i~k}{8\pi} \int_{AdS_3} AdA - \frac{k}{16 \pi}\ \int_{\d AdS_3} d\tu dy\ 
\sqrt{h}\ h^{\a\b}A_\a A_\b -\f {i k}{4\pi} \int_{AdS_3}{d\tau dz dy\ \sqrt{g^{(3)}}~ A_M J^M}.  
\end{equation}
The equations of motion from above action are
\begin{subequations}
 \begin{align}
F_{zy} &= 0, \\
F_{y \tau } &= 0,\\
F_{\tau z} &= \s{g^{(3)}}\ J_0.
 \end{align}
\end{subequations}
Let us consider the solution of the metric to be 
\begin{eqnarray}
 ds_{\rm{3DBTZ}}^2 = \frac{l^2}{z^2}\left(\frac{M z^2}{l}-1\right)^2 d\tu^2 +\frac{l^2}{z^2}dz^2+\frac{ a^2}{z^2}\left(\frac{M z^2}{l}+1\right)^2 dy^2
\end{eqnarray}  
here $y \in \{0,2 \pi l\}$ and $\tau \in \{0,\beta \}$.
The solution to above equations in the $A_z =0$ gauge and with above metric is
\begin{subequations}
 \begin{align}
  A_{\tau}(\tau,z) &= \f{J_0 a l^2}{2} \left(\f{1}{z^2} + \f{\pi^4}{\beta^4}\ z^2 - \f{2\pi^2}{\beta^2}\right),\\
  A_y(\tau,z) &= constant.
 \end{align}
\end{subequations}
Here we have used $(\pi/\beta)^2 = M/l$ and also noted that the integration constant in the solution for $A_\tau$ is fixed by demanding that the field vanishes at the horizon, $z =\beta / \pi$.
Plugging these solutions into the action, we get 
\begin{equation}
 S_{\text{3D-Gauge}}^{^{on-shell}} \approx - \f{k\beta l}{8}\ \left[ \f{a^3 J^2_0 l^3}{4 \delta^4} - \f{a^3 J^2_0 l^3 \pi^2}{2 \beta^2 \delta^2} + \f{A_y^2 l}{a} + O(\delta^2)\right] 
\end{equation}
here we have chosen the boundary surface at $z = \delta$. In the limit $\delta \to 0$ the on-shell action is divergent. We find that these divergences are cured by the following covariant local counter terms:
\begin{equation}\label{3dct}
 S_{\text{3D-CT}} = \f{k l}{64 \pi}\int A \wedge J + \f{k}{16 \pi l} \int \f{(A^\mu A_\mu)}{(J^\mu J_\mu)}\ (A \wedge J)
\end{equation}
The evaluation of these terms on the above solutions gives
\begin{equation}
  S_{\text{3D-CT}}^{^{on-shell}} \approx  \f{k\beta l}{8}\ \left[ \f{a^3 J^2_0 l^3}{4 \delta^4} - \f{a^3 J^2_0 l^3 \pi^2}{2 \beta^2 \delta^2} + \f{a^3 J^2_0 l^3 \pi^4}{2 \beta^4 } + O(\delta^2)\right],
\end{equation}
hence 
\begin{equation}
 S_{\text{3DGauge}}^{^{on-shell}} +   S_{\rm{3D-CT}}^{^{on-shell}}  = \f{k \beta l}{8} \left( \f{a^3 J^2_0 l^3 \pi^4}{2 \beta^4} - \f{A_y^2 l}{a}\right) + O\left(\delta^2\right)
\end{equation}
which is divergence free.\\

A  KK reduction to 2D is performed by writing  
\begin{align}
 A &= A_\tau\ d\tau + \rchi \ dy, \nn\\
ds^2&=g^{(3)}dx^M dx^N = g^{(2)}_{\mu \nu } dx^\mu dx^\nu + \Phi^2 (dy + l B_\mu dx^\mu)^2.\nn\\
h^{\tau \tau} &= \gamma^{\tau \tau},\nn \\
h_{yy} &= \Phi^2
\end{align}
hence \eq{3dct} reduces to
\begin{align}
  S_{\rm{3D-CT}} &= \f{k l}{64 \pi}\int A \wedge J + \f{k}{16 \pi l} \int \f{(A^\mu A_\mu)}{(J^\mu J_\mu)}\ (A \wedge J) \nn \\
   &= \f{k l}{64 \pi}\int d\tau dy\ J_0\ A_\tau\ h_{yy}  + \f{k}{16 \pi l} \int d\tau dy\ \f{h^{\tau \tau}\ A_\tau^3 }{J_0} \nn\\
   & =  \f{k  J_0l^2}{32} \int d\tau\ \Phi^2\ A_\tau\ + \f{k}{8 J_0} \int d\tau\ A_\tau\ (\gamma^{\tau \tau}\ A_\tau^2)
\end{align}
The last line in the above action contains exactly the counter terms proposed in \eq{eq:ct} for the 2D theory.

\subsection{Cancellation of divergences with the boundary counter terms}
In this section we will show that the counter terms, \eq{eq:ct} proposed in section \ref{sec:ct} cancel the divergence coming from the first term in \eq{eq:total-delta-s} (we will denote this term in this section by $S_{boundary}$). Let us first evaluate this term using \eq{gammauu}, \eq{phir} and \eq{Aufin}.
\begin{align}
 S_{boundary} &= -\f{k l}{8} \int_{\del M_2} d u ~\Phi\ \sqrt{\gamma}\ \gamma^{u u} A_u A_u \nn \\
  &= -\int_{\del {M_2}} du \left(\frac{J_0^2 k l^4 \Phi_r(u)^3}{32 \delta^4} + \frac{J_0 k l^2 \Phi_r(u)^2 (B\ \varphi'(u))}{8 \delta^2} + \frac{k \Phi_r(u)}{8}  (B\   \varphi'(u))^2 + O\left(\delta ^2\right) \right).
\end{align}
Now let us evaluate the counter terms
\begin{equation}
S_{CT} = S^1_{CT} + S^2_{CT} = \f{k J_0 l^2}{32} \int_{\del M_2} du\ \Phi^2 A_u + \f{k}{8J_0} \int_{\del M_2} du\ A_u (\gamma^{uu} A_u A_u).  
\end{equation}
The first term in the above expression is
\begin{align}
  S^1_{CT}&=\f{k J_0 l^2}{32} \int_{\del M_2} du\ \Phi^2 A_u  \hspace{13cm}\nn \\
 &= \int_{\del M_2} du \left( \frac{J_0^2 k l^4 \Phi_r(u)^3}{64 \delta ^4}+\frac{J_0 k l^2 \Phi_r(u)^2 (B\  \varphi'(u))}{32 \delta ^2}+O\left(\delta ^2\right) \right) \hspace{6cm}
\end{align}
and the second term evaluates to
\begin{align}
 S^2_{CT} &= \f{k}{8J_0} \int_{\del M_2} du\ (\gamma^{uu}\ A_u A_u )\ A_u \hspace{10cm}\nn \\
 &=  \int_{\del M_2} du \left(\frac{J_0^2 k l^4 \Phi_r(u)^3}{64 \delta ^4}+\frac{3 J_0 k l^2 \Phi_r(u)^2 (B\  \varphi'(u))}{32 \delta ^2}+\frac{3k \Phi_r(u)}{16}  (B\  \varphi'(u))^2+O\left(\delta ^2\right)\right).
\end{align}
It is easy to see that the combination
\begin{equation}
S_{boundary} + S_{CT} = \f{k}{16}\ \int_{\del M_2} du\ \Phi_r(u)\ (B\ \varphi'(u))^2
\end{equation}
does not diverge as $\delta \to 0$ and we have a renormalized on-shell action.

We can also consider a gauge transformed gauge field, $A^\phi_u$ given by
\begin{equation}
 A_u \to A^\phi_u = A_u + \partial_u\phi(u)
\end{equation}
to evaluate the combination $S_{boundary} + S_{CT}$ and the result is
\begin{equation}
 S_{boundary} + S_{CT} = \f{k}{16}\ \int_{\del M_2} du\ \Phi_r(u)\ (\phi'(u)+ B\ \varphi'(u))^2
\end{equation}
which is also finite as $\delta \to 0$. This result will be used to derive \eq{sigma-model}.

\section{Quantum chaos from  minimally coupled probe scalars in 3D
  \label{app:chaos-3D}}

In this section, we will discuss charged scalars coupled to the
dilaton and the metric in 2D, given by the action
\begin{align}
S_{matter} = - \f1{16\pi G_2}  \int d^2x \sqrt{g}\ \Phi
\left(g^{\al\be} D_\al \eta^* D_\be \eta +  m^2 |\eta|^2\right)
\label{charged-matter-dilaton}
\end{align}
in stead of the action \eq{charged-matter}. The present action is more
natural from the 3D point of view, as it comes from a a minimally
coupled scalar in 3D:
\begin{equation}
  S_3 = -\f1{16\pi G_3}
  \int d^3 x \s{g^{(3)}} \left( g^{(3)}_{MN}\ \del^M \eta^*\ \del^N \eta +
  m^2 |\eta|^2\right),
  \label{3D-action}
\end{equation}
As mentioned in the text, we have taken here the 3D Euclidean metric
to be of the form ($x^M = (x^\mu, y),\; x^\mu=(\tau,z)$),
\begin{equation}
 ds^2=g^{(3)}_{MN}dx^M dx^N= g_{\mu \nu }(z) dx^\mu dx^\nu + \Phi(z)^2 dy^2
\end{equation}
where we take both the $y$ and $\tau$ circles to be compact, with $y \in
(0, 2\pi l)$ and $\tau \in (0, \beta)$.

\paragraph{Charge-neutral case}
To begin, let us consider the pure gravity situation, relevant for the
uncharged SYK model. In \eq{charged-matter-dilaton}, we need to
consider $\eta$ to be real. We claim that, if the scalar $\eta$ is governed
by this new action, with coupling to the dilaton, then the
discussion in Section \ref{sec:neutral-OTO} remains fairly unchanged
except that  the mass-dimension relation
\eq{mass-dim-2D} gets changed to
\begin{align}
  \Delta = 1/2 + \nu_2, \; \nu_2= \sqrt{1 + m^2 l^2}.
  \label{mass-dim-hybrid}
\end{align}
Essentially the quantity $\nu$ in \eq{mass-dim-2D} changes to
\[
\nu \to \nu_2= \sqrt{1 + m^2 l^2}
\]
This modification does not affect the discussion of the Liapunov
exponent in Section \ref{sec:neutral-OTO}. Note that the new dimension
formula is a kind of hybrid of 2D and 3D bulk; the origin
of this will be clear below.

The way we prove the new mass-dimension formula is the following.  For
illustrative purposes, we first start with the AdS-Poincare metric in
3D ($\beta=\infty$). If we applied the usual machinery of AdS/CFT to
the case of the 3D bulk, before doing any KK reduction, we would
obtain, after the appropriate renormalizations \cite{deHaro:2000vlm},
that
\begin{align}
  S_{\text{3D-onshell}}= \f{1}{2}\int dk_y \ dk_\tau\  \eta_{ren}(K)\ \eta_{ren}(-K)\  |K|^{2\nu_2},
  \quad \nu_2= \sqrt{1+ m^2 l^2}
  \label{3D-onshell}
\end{align}
Here $K=(k_y, k_\tau)$ are the momenta along $y$ and $\tau$ respectively;
$\eta_{ren} \sim \delta^{-(1-\nu_2)}\, \eta_{boundary}$, where the 2D
boundary is placed at $z=\delta$.  In 3D bulk (equivalently, for 2D
CFT), before doing any KK reduction, we would identify the coefficient
of the quanratic $\eta$ term as two-point function of the operator $O$
in momentum space, so that in real space $X=(y,\tau)$ it would be
\begin{align}
  \lan O(X_1) O(X_2) \ran = \int dk_y\ dk_\tau\ |K|^{2\nu_2} \exp[i K.X] \propto
  |X|^{- 2 \Delta_2}, \quad X=X_1 - X_2, \; \Delta_2 = 1 + \nu_2
  \label{2D-fourier}
\end{align}
When we perform a KK reduction, the computation \eq{3D-onshell},
of course, remains valid; however the last computation \eq{2D-fourier},
is replaced by a one-dimensional Fourier transform (since
KK reduction implies $k_y=0$):
\begin{align}
  \lan O(\tau_1) O(\tau_2) \ran = \int dk_\tau\ |k_\tau|^{2\nu_3}\ \exp[i k_\tau \tau ] \propto
  |\tau|^{- 2 \Delta}, \quad \tau \equiv \tau_1 - \tau_2,\; \Delta = 1/2 + \nu_3.
  \label{2D-fourier}
\end{align}
So the operator dimension $\Delta$, corresponding to a mass $m$ of a
neutral scalar coupled to the dilaton and the metric, as in
\eq{charged-matter-dilaton} with real $\eta$, is given by the new
mass-dimension formula \eq{mass-dim-hybrid}, as claimed above. In case
of finite temperature, by following the usual steps (such as (a)
summing over images, to account for the periodic
identification of the thermal circle, or (b) replacing the Fourier
transform in the above equation by a Matsubara sum) we find that the
boundary correlator of the dual operator $O$ is given by the equation
\eq{thermal-2-pt} with $\Delta$ given by the new formula
\eq{charged-matter-dilaton}.

\paragraph{Charged case}

It is easy to show that the discussion of quantum chaos in this case
too, for a charged scalar coupled to the dilaton and the metric as in
\eq{charged-matter-dilaton}, does not have any other modification,
except for the mass-dimension formula which is again given by
\eq{mass-dim-hybrid}. Once again, this does not affect any of the
conclusion arrived at in the main text.

\paragraph{Summary} In summary, we have shown above that if the scalar
field is coupled to the dilaton as well as to the metric, as is
natural from the 3D viewpoint, the mass of the scalar field, dual to a
given operator $O$ of dimension $\Delta$, is to be chosen according to
the new mass-dimension formula \eq{mass-dim-hybrid}, in stead of the standard
mass-dimension formula \eq{mass-dim-2D}. 

\end{appendices}

\bibliographystyle{JHEP} 
\bibliography{1g}

\end{document}